\documentclass[journal,12pt, draftclsnofoot,onecolumn]{IEEEtran}
\IEEEoverridecommandlockouts
\usepackage{setspace}
\usepackage{cite}
\usepackage{amsmath,amssymb,amsfonts,mathtools,bm}
\usepackage{amsthm}
\usepackage{algorithm}
\usepackage{graphicx}
\usepackage{textcomp}
\usepackage{xcolor,float}
\usepackage{tabularx}
\usepackage{textcomp}
\usepackage{diagbox}
\usepackage{svg}
\usepackage{booktabs}
\usepackage{subfigure}
\usepackage{hyperref}
\newtheorem{theorem}{Theorem}
\newtheorem{problem}{Problem}
\usepackage{algpseudocode}
\usepackage{graphicx}
\usepackage{makecell}
\usepackage{amsthm}
\usepackage{caption}
\usepackage{sidecap}
\usepackage{microtype}
\usepackage{booktabs} 
\usepackage{multirow}
\usepackage{float}
\usepackage{adjustbox} 
\usepackage{amsmath}
\usepackage{amssymb}
\usepackage{colortbl}
\usepackage{makecell}
\usepackage{float}
\usepackage{url}
\usepackage{balance}
\usepackage{bbding}
\usepackage{hyperref}
\usepackage{graphicx}
\usepackage{sidecap}
\usepackage{microtype}
\usepackage{graphicx}
\usepackage{subfigure}
\usepackage{booktabs} 
\usepackage{multirow}
\usepackage{array}
\usepackage{subcaption}
\usepackage{graphicx}
\usepackage{makecell}
\usepackage{amsmath}
\usepackage{amssymb}

\newtheorem{remark}{Remark}
\usepackage{tikz}

\allowdisplaybreaks
\renewcommand{\baselinestretch}{1}

\begin{document}
\doublespacing

\title{Erasing Noise in Signal Detection with Diffusion Model: From Theory to Application}
    
\author{
Xiucheng Wang,
Peilin Zheng,
Nan Cheng
\thanks{ }
\thanks{
\par Xiucheng Wang, Peilin Zheng, Nan Cheng are with the State Key Laboratory of ISN and School of Telecommunications Engineering, Xidian University, Xi’an 710071, China.
}

}

\maketitle
\IEEEdisplaynontitleabstractindextext

\IEEEpeerreviewmaketitle

\begin{abstract}
In this paper, a signal detection method based on the denoise diffusion model (DM) is proposed, which outperforms the maximum likelihood (ML) estimation method that has long been regarded as the optimal signal detection technique. Theoretically, a novel mathematical theory for intelligent signal detection based on stochastic differential equations (SDEs) is established in this paper, demonstrating the effectiveness of DM in reducing the additive white Gaussian noise in received signals. Moreover, a mathematical relationship between the signal-to-noise ratio (SNR) and the timestep in DM is established, revealing that for any given SNR, a corresponding optimal timestep can be identified. Furthermore, to address potential issues with out-of-distribution inputs in the DM, we employ a mathematical scaling technique that allows the trained DM to handle signal detection across a wide range of SNRs without any fine-tuning. Building on the above theoretical foundation, we propose a DM-based signal detection method, with the diffusion transformer (DiT) serving as the backbone neural network, whose computational complexity of this method is $\mathcal{O}(n^2)$. Simulation results demonstrate that, for BPSK and QAM modulation schemes, the DM-based method achieves a significantly lower symbol error rate (SER) compared to ML estimation, while maintaining a much lower computational complexity.
\end{abstract}


\section{Introduction}
Signal detection plays a critical role in digital baseband transmission, since it estimates which symbols are transmitted by the sender, from the noisy received signals. Thus, the performance of signal detection directly impacts the symbol error rate (SER) of data transmission, which in turn determines the error-free transmission rate, also known as the Shannon threshold \cite{shannon1948mathematical}. As a result, numerous signal detection techniques have been developed to minimize the SER and bring the transmission rate as close as possible to the Shannon threshold. However, signal detection is a non-deterministic polynomial hard (NP-hard) problem since transmitted symbols typically have discrete phases or amplitudes, and they are uncorrelated with one another \cite{zhang2022efficient}. Consequently, conventional signal detection algorithms often rely on linear methods to offer a low computational complexity, such as minimum mean square error (MMSE), matched filtering (MF), and zero-forcing (ZF) \cite{4749305,9788039}. Despite high computing efficiency, linear detection methods have a significant performance gap compared to optimal signal detection techniques. Meanwhile, as a classical non-linear signal detection method, the maximum likelihood (ML) estimation method is widely regarded as the optimal signal detection approach, since it exhaustively enumerates all possible symbol combinations and selects the one with the highest likelihood as the estimated transmission symbols \cite{dorfman1969maximum}. Although ML estimation has an impressive low SER, its computational complexity, $\mathcal{O}(2^n)$ \cite{zhang2022efficient}, is prohibitively high, particularly for large-scale systems. Therefore, methods designed based on conventional signal detection theory not only exhibit lower efficiency, but their performance upper bounds are also unable to outperform ML estimations.

To address the challenge of balancing low computational complexity with low SER in traditional signal detection methods, researchers have explored the use of inference-efficient neural networks (NNs) for signal detection. One pioneering example is the DnCNN-based signal detection method \cite{dncnn}, which adapts the DnCNN, originally designed for image denoising in computer vision, to decrease the influence of noise on the received symbols for signal detection. However, as DnCNN is built upon the convolutional neural network (CNN) architecture, it performs well only when the transmitted symbols are correlated \cite{zhang2017beyond}. Its performance degrades significantly when the symbols are randomized. To overcome this limitation, the ComNet is proposed in \cite{comnet}, which is a hybrid model-based signal detection approach that combines traditional expert knowledge in signal detection with the feature extraction capabilities of multilayer perceptrons (MLPs). In recent years, inspired by advancements in Transformers, an end-to-end signal detection method based on Transformer architecture is introduced in the \cite{sigt}. Moreover, \cite{message-passing} leverages the powerful message-passing capabilities of the graph neural network (GNN) in multiple-input multiple-output (MIMO) systems and utilizing the nonlinear structure-fitting capacity of NNs, to further improve the performance of signal detection.

However, all existing NN-based signal detection methods face two significant challenges: poor generalization across different signal-to-noise ratios (SNRs) and a huge performance gap compared to traditional optimal signal detection methods. Unfortunately, these two challenges are caused by the inherent features of the traditional NN training methods. According to the universal approximation theorem, a trained NN can extract and fit a specific form of data distribution \cite{goodfellow2016deep}. However, in signal detection, the distributions of received signals exhibit significant variations under different SNRs due to the variety in noise intensity. Consequently, training a single NN to effectively extract features from received signals across a wide range of SNRs poses a considerable challenge when employing traditional learning methods. This limitation arises from the NN's inherent difficulty in generalizing across diverse noise conditions, potentially leading to poor performance in practical scenarios where SNR variability is common. Current NN-based methods typically train on datasets with a specific SNR and test the NN performance with a similar SNR, which results in a dramatic degradation in performance when these models encounter received signals at different SNRs from the training data. This occurs because the input data is out-of-distribution (OOD) relative to the training data \cite{hsu2020generalized}. To reduce the impact of OOD data on NN-based signal detection, fine-tuning or transformations of data distribution features are often required. Despite these efforts, NN-based methods still always exhibit a large performance gap when compared to optimal ML estimation. This gap is prevalent in merely all NN-based signal detection methods, no matter which NN architecture is used, such as  MLP, CNN, and more advanced architectures like Transformers and GNNs. Therefore, simply increasing the structural complexity or the number of parameters in NNs is even insufficient to bridge this performance gap with traditional non-linear signal detection, let alone outperform the ML.

Therefore, it is the corner to rethink why NN-based methods have poor generalizability to various SNRs and can NN-based methods outperform the traditional optimal signal detection method. The core issue is the absence of a novel theoretical framework for intelligent signal detection that extends beyond traditional signal detection theories. To establish such an intelligent signal detection theory, it is essential to identify which factor constrains the performance of traditional optimal signal detection methods. The key limiting factor is the \textbf{additive noise} in the received signal. Since the traditional signal detection theory has a non-proved assumption, that the intensity of additive noise in the received signal is equivalent to the noise intensity in the channel through which it has passed. Therefore, if we can theoretically prove that this noise can be effectively erased or reduced, a new signal detection method that is beyond classical ML estimation could be developed. Fortunately, the denoise diffusion model (DM) \cite{ho2020denoising}, though widely applied in the field of content generation, offers a promising solution by canceling Gaussian noise in data, since in mathematics the procedure of content generation in DM is the procedure of reducing the Gaussian noise adding on the generated content. Since the success in content generation applications of DM has empirically shown that the DM can significantly reduce the noise added to the generated content, it is also promising to reduce the noise added to the received signals, allowing for signal detection performance that exceeds that of ML estimation. By reformulating the classical discrete denoise diffusion probability model into continuous stochastic differential equations (SDEs), we demonstrate that for any SNR, there exists a corresponding timestep in the DM that can be used as the start point to reduce the noise on the receiving signal, which also addresses the generalization issues of NN-based methods for different SNRs. The main contributions of this paper are as follows.
\begin{enumerate}
    \item A novel intelligent signal detection theory that goes beyond traditional signal detection theory is established in this paper. By employing stochastic differential equations (SDEs), we theoretically prove that the additive Gaussian white noise in received signals can be effectively reduced using the denoise diffusion model. Therefore, in theory, the performance of the classical optimal signal detection method that is ML estimation, can be exceeded.
    \item A mathematical relationship between the SNR of an arbitrary received signal and the timestep in DM through theoretical analysis is established. The further analysis also reveals that the challenge of data distribution mismatch between the received signal to be detected and the training data can be resolved by appropriately scaling the received signal. As a result, an NN trained under our proposed intelligent signal detection theory framework can be utilized for detection across a wide range of SNRs without the need for fine-tuning.
    \item Based on the established new intelligent signal detection theory, a DM-based signal detection framework is proposed, where only one step of linear operation is needed to detect the signal after the noise in the received signal is reduced by the DM. Specifically, to further enhance signal detection performance, a diffusion transformer (DiT) is the backbone NN in DM, where we patchify the received signals into multiple tokens as the input of DiT, thus a low computational complexity of $\mathcal{O}(n^2)$ can be obtained.
    \item Simulation results demonstrate that the proposed DM-based signal detection method for BPSK and 4QAM modulation schemes outperforms the ML estimation method. This represents a significant breakthrough in traditional signal detection theory.
\end{enumerate}

\section{Preliminaries}\label{sec2}
Diffusion models are a class of generative models based on Markov chains that progressively restore data through a learned denoising process. These models have emerged as strong competitors to generative adversarial networks (GANs) in various generative tasks, such as computer vision \cite{lin2023diffbir} and natural language processing \cite{austin2021structured}. Moreover, diffusion models exhibit significant potential in perception tasks, including image segmentation, object detection, and model-based reinforcement learning (RL) \cite{LDM}. 

In diffusion models, there are two main procedures: the forward diffusion procedure, where raw data are progressively diffused into noise, and the reverse denoising procedure, where a NN is employed to remove the noise added to the data, thus generating raw data from noise. From a probabilistic modeling standpoint, the essence of generative models lies in training them to produce data $\hat{\bm{x}} \sim p_\theta(\hat{\bm{x}})$ that mirror the distribution of the training data $\bm{x} \sim p_{\text{data}}(\bm{x})$. The denoising diffusion probabilistic model (DDPM) employs two Markov chains: a forward chain that converts data into noise and a backward chain that reconstructs the data from noise. Given the data $\bm{x}_0$, the progression of a forward Markov chain is realized by generating a series of stochastic variables $\bm{x}_1, \bm{x}_2, \ldots, \bm{x}_T$, which evolve following the transition kernel $q(\bm{x}_t \mid \bm{x}_{t-1})$. By employing the chain rule of probability in conjunction with the Markov property, the joint probability distribution of $\bm{x}_1, \ldots, \bm{x}_T$ given $\bm{x}_0$ can be factorized as
\begin{equation}
q(\bm{x}_1, \ldots, \bm{x}_T \mid \bm{x}_0) = \prod_{t=1}^T q(\bm{x}_t \mid \bm{x}_{t-1}).
\end{equation}
Thus, the forward noising process, which produces latent variables $\bm{x}_t$ from $\bm{x}_0$ by adding Gaussian noise at each time step $t \in \{1, 2, \ldots, T\}$, is defined as
\begin{equation}
q\left(\bm{x}_t \mid \bm{x}_{t-1}\right) = \mathcal{N}\left(\sqrt{1 - \beta_t} \, \bm{x}_{t-1}, \beta_t \bm{I}\right),
\end{equation}
where $T$ is the total number of diffusion iterations, $\beta_t \in (0,1)$ is a hyperparameter controlling the variance scaling factor, $\mathcal{N}(\cdot, \cdot)$ denotes the Gaussian distribution, and $\bm{I}$ is the identity matrix. By setting $\alpha_t = 1 - \beta_t$ and $\bar{\alpha}_t = \prod_{s=1}^{t} \alpha_s$, the distribution of $\bm{x}_{t}$ conditioned on $\bm{x}_0$ can be obtained as
\begin{equation}
q(\bm{x}_t \mid \bm{x}_0) = \mathcal{N}\left(\sqrt{\bar{\alpha}_t} \, \bm{x}_0, (1 - \bar{\alpha}_t) \bm{I}\right), \label{diffused}
\end{equation}
which implies
\begin{equation}
\bm{x}_t = \sqrt{\bar{\alpha}_t} \, \bm{x}_0 + \sqrt{1 - \bar{\alpha}_t} \, \bm{\epsilon},
\end{equation}
where $\bm{\epsilon} \sim \mathcal{N}(\bm{0}, \bm{I})$.

For data generation, the DDPM initially creates unstructured noise vectors from the prior distribution, subsequently removing the noise through a learnable Markov chain operated in reverse temporal order. The reverse process is formulated as
\begin{equation}
p_{\bm{\theta}}(\bm{x}_{t-1} \mid \bm{x}_t) = \mathcal{N}\left(\bm{\mu}_{\bm{\theta}}(\bm{x}_t, t), \beta_t \bm{I}\right), \label{reverse-2}
\end{equation}
where $\bm{\mu}_{\bm{\theta}}(\bm{x}_t, t)$ is a neural network with trainable parameters $\bm{\theta}$, and $\beta_t$ is the same hyperparameter as in the forward process. According to \cite{LDM}, by applying the trained neural network $\bm{\mu}_{\bm{\theta}}$, we iteratively denoise $\bm{x}_t$ from $t = T$ down to $t = 1$ as
\begin{equation}
\bm{x}_{t-1} = \frac{1}{\sqrt{\alpha_t}} \left( \bm{x}_t - \frac{1 - \alpha_t}{\sqrt{1 - \bar{\alpha}_t}} \bm{\mu}_{\bm{\theta}}(\bm{x}_t, t) \right) + \beta_t \bm{z}, \label{ddpm-reverse}
\end{equation}
where $\bm{z} \sim \mathcal{N}(\bm{0}, \bm{I})$. Notably, the addition of the noise term $\beta_t \bm{z}$ in \eqref{ddpm-reverse} ensures that the distribution of $\bm{x}_{t-1}$ obtained through denoising matches, particularly in terms of variance, the distribution derived from the forward diffusion process. If this noise term is omitted, the denoised $\bm{x}_{t-1}$ corresponds to the mean of the distribution obtained via the forward diffusion procedure. Strictly speaking, the scaling factor in the noise term of \eqref{ddpm-reverse} should be $\tilde{\beta}_t = \frac{1 - \bar{\alpha}_{t-1}}{1 - \bar{\alpha}_t} \beta_t$ to ensure distribution consistency. However, as shown in \cite{ho2020denoising}, setting the scaling factor directly to $\beta_t$ achieves comparable performance while reducing computational complexity. Therefore, the scaling factor in the noise term of \eqref{ddpm-reverse} is set to $\beta_t$ for both efficiency and effectiveness.

\begin{table}
    \centering
    \caption{Notation Table}
    \resizebox{0.7\linewidth}{!}{
    \begin{tabular}{c|c}
    \hline
       Variables  & Definition\\ \hline
       $\mathbb{E}[\cdot]$ & The expectation of random variables.\\
       $\mathbb{D}[\cdot]$ & The variance of random variables.\\
       $|\cdot|$ & The cardinality of the vector or matrix.\\
        $\bm{s}$ & The transmitted signal symbol.\\
        $\bm{r}$ & The received signal.\\
        $\bm{n}$ & The noise added to the transmitted signal symbol.\\
        $\sigma^2$ & The noise power in the received signal.\\
        $\bm{H}$ & The channel matrix.\\
        $t$ & The timestep in the diffusion process.\\
        $\bm{x}_{t}$ & Noise data after $t$ times of diffusion. \\ 
        $\bm{\epsilon}_{t}$ & The noise in the diffusion procedure.\\
        $\bm{h}_{t}$ & The latent representation of $\bm{x}_{t}$.\\
        $\alpha$ & The scaling factor.\\
        \hline
       \end{tabular}
    }
    \label{tab-notation}
\end{table}

\section{System Model and Problem Formulation}
In this paper, we address a fundamental signal detection problem: how to accurately recover the original transmitted signal from a noisy received signal after it passes through a noisy channel. Let the transmitted signal in the time domain be denoted by $\bm{s}$, which consists of multiple symbols. For generality, we assume that the transmitted signal contains no semantic information, and the symbols are independent of one another. Specifically, each symbol in $\bm{s}$ is randomly selected from the symbol set $\mathcal{S} = {\bm{s}_1, \cdots, \bm{s}_m}$ with equal probability, and the duration of each symbol is $T_s$. Additionally, we allow for varying symbol energies, i.e., $|\bm{s}_{i}|^{2}$ may differ for different $i \in \{1, \dots, |\mathcal{S}|\}$, adding flexibility to the problem. The signal is transmitted through a noisy channel $\bm{H}\in \mathbb{R}^{N_r\times N_r}$, where $N_r$ represents the number of transmission antennas, and the noise has a variance of $\sigma^2$. The received signal is denoted by $\bm{r}$. The objective is to design a decision rule $\bm{\varphi}(\cdot)$ that maximizes the probability of correctly identifying the transmitted symbols based on the noisy observation $\bm{r}$. Traditionally, the above problem can be formally described as follows.
\begin{problem}\label{p1}
\begin{align}
    &\max_{\bm{\varphi}(\cdot)}\;\; \mathbb{E}_{\bm{r}\sim q(\bm{r})}\left[\mathbb{I}\left(\bm{\varphi}(\bm{r},\mathcal{S})=\bm{s}\right)\right],\label{obj-1}\\
    &s.t.\quad\;\, \bm{r}=\bm{H}\bm{s}+\bm{n},\label{c1-1}\tag{\ref{obj-1}a}\\
    &\qquad\;\;\, \mathbb{E}\left[\bm{n}\right]=0,\label{c1-2}\tag{\ref{obj-1}b}\\
    &\qquad\;\;\, \mathbb{E}\left[\|\bm{n}\|^{2}\right]=\hat{\sigma}^2,\label{c1-3}\tag{\ref{obj-1}c}
\end{align}
\end{problem}
\noindent where $\hat{\sigma}^2$ is the noise density of the added to the received signal symbol $\bm{r}$, which is traditionally assumed the same as the density of noise $\sigma^2$ in the channel, and $q(\bm{r})$ is the probability distribution of the received signal $\bm{r}$, and $\mathbb{I}(\cdot)$ is the indicator function which is equal to 1 if the input is true, otherwise is 0. The objective in \eqref{obj-1} is to maximize the probability that the transmission symbol $\bm{s}$ is correctly received by the decision rule $\bm{\varphi}(\bm{r},\mathcal{S})$ from the received noisy signal $\bm{r}$. \eqref{c1-2} and \eqref{c1-3} describe the received signal $\bm{r}$ as the sum of the transmitted signal $\bm{s}$ and additive noise $\bm{n}$, where $\bm{n}$ is normally distributed with zero mean and covariance $\hat{\sigma}^2 \bm{I}$.

\begin{figure}
    \centering
    \includegraphics[width=1\linewidth]{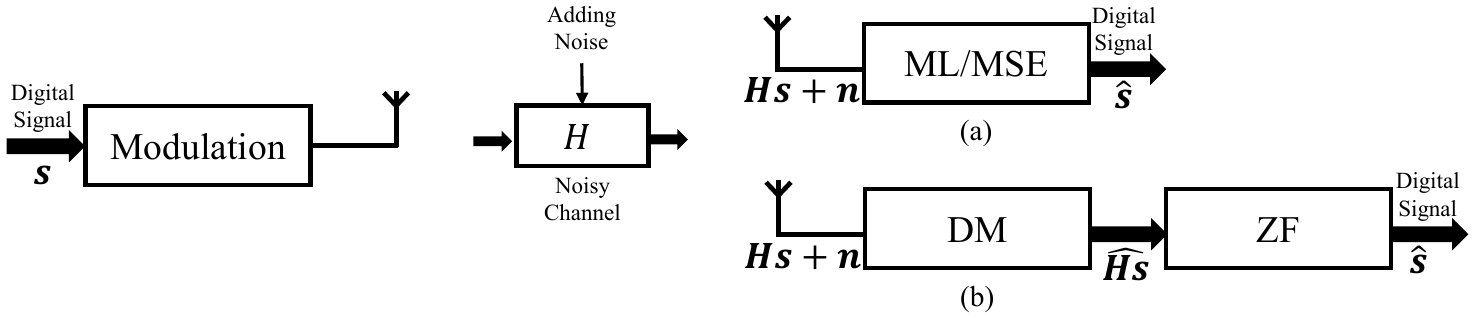}
    \caption{Block diagram of the signal detection method, (a) is the traditional signal detection system, and (b) is the proposed DM-based signal detection system.}
    \label{fig-1}
\end{figure}
Traditional signal detection theory assumes that in Problem~\ref{p1} the noise intensity in the signal to be received is equal to the noise intensity of the channel through which the signal passes. This assumption inherently limits the performance of traditional signal detection methods. Therefore, if there exists a method to reduce the noise intensity in the received signal such that $\hat{\sigma}^2 < \sigma^2$, we can effectively decrease the equivalent noise in the received signal. This reduction allows us to achieve a lower probability of misjudgment, thereby surpassing the performance limits set by traditional signal detection theory. Suppose there exists a transformation $\bm{\pi}(\cdot)$ that yields $\hat{\sigma} < \sigma$. In that case, we can reformulate Problem~\ref{p1} to incorporate this reduced noise intensity, potentially leading to improved signal detection performance.

\begin{problem}\label{p2}
\begin{align}
    &\max_{\bm{\varphi}(\cdot),\bm{\pi}(\cdot)}\;\; \mathbb{E}_{\bm{r}\sim q(\bm{r})}\left[\mathbb{I}\left(\bm{\varphi}(\bm{r},\mathcal{S})=\bm{s}\right)\right],\label{obj-2}\\
    &s.t.\quad\;\;\;\,\,\, \hat{\bm{r}}=\bm{\pi}(\bm{r}),\label{c2-2}\tag{\ref{obj-2}b}\\
    &\qquad\;\;\;\;\,\,\, \bm{r}=\bm{H}\bm{s}+\bm{n},\label{c2-3}\tag{\ref{obj-2}c}\\
    &\qquad\;\;\;\;\,\,\, \mathbb{E}\left[\bm{n}\right]=0,\label{c1-3}\tag{\ref{obj-2}d}\\
    &\qquad\;\;\;\;\,\,\, \mathbb{E}\left[\|\bm{n}\|^{2}\right]=\sigma^2,\label{c1-4}\tag{\ref{obj-2}e}
\end{align}
\end{problem}

Obviously, Problem~\ref{p2} can be decomposed into two independent sub-problems connected in stages, namely, reducing the noise density in the received symbol $\bm{r}$ and judging it according to the denoised symbol $\hat{\bm{r}}$. Therefore, in order to solve Problem~\ref{p2}, this paper mainly focuses on how to reduce noise density in the received symbol $\bm{r}$. The problem of reducing the noise density in the received symbol is equal to increasing the SNR of the received signal, which can be formulated as follows.
\begin{problem}\label{p3}
    \begin{align}
        &\max_{\bm{\pi}(\cdot)}\;\;\mathbb{E}_{\bm{r}\sim q(\bm{r})} \left[\frac{\|\bm{Hs}\|^2}{\|\hat{\bm{n}}\|^2}\right],\label{obj-3}\\
        &s.t.\quad\;\, \hat{\bm{n}}=\hat{\bm{r}}-\bm{H}\bm{s},\label{c3-1}\tag{\ref{obj-3}a}\\
        &\qquad\;\;\, \hat{\bm{r}} = \bm{\pi}(\bm{r}),\label{c3-2}\tag{\ref{obj-3}b}\\
        &\qquad\;\;\, \bm{r}=\bm{Hs}+\bm{n},\label{c3-3}\tag{\ref{obj-3}c}\\
        &\qquad\;\;\, \mathbb{E}\left[\bm{n}\right]=0,\label{c1-3}\tag{\ref{obj-3}d}\\
    &\qquad\;\;\, \mathbb{E}\left[\|\bm{n}\|^{2}\right]=\sigma^2,\label{c1-4}\tag{\ref{obj-3}e}
    \end{align}
\end{problem}

\section{DM-based Intelligent signal Transmission Theory}\label{sec4}
Since the additive white Gaussian noise (AWGN) channel is the most fundamental and representative channel model in digital communication theory, as established by Shannon in his seminal work \cite{shannon1948mathematical}, we begin our study by using the AWGN channel to explore how DMs can achieve efficient signal detection. In this section, we first demonstrate that the SDE-based diffusion process can effectively remove Gaussian noise from any data, as detailed in Section~\ref{sec4-1}. Subsequently, in Section~\ref{sec4-2}, we establish the correlation between the diffusion process and the denoising of received signals.
\subsection{SDE of Diffusion Procedure}\label{sec4-1}
According to \cite{song2020score}, the Markov process formulated of diffusion process in Section \ref{sec2} can also be expressed using an SDE as follows.
\begin{align}
d\bm{x}_t = f_t \bm{x}_t dt + g_t d\bm{w}_t,
\end{align}
where $f_t = \frac{d \log \alpha_t}{dt}$, $g_t^2 = \frac{d \beta_t^2}{dt} - 2 f_t \beta_t^2$, and $\bm{w}_t$ is a standard Wiener process. By letting $f_t \bm{x}_t = \bm{h}_{t}$ and $g_t = 1$, an integral form of the SDE can be obtained as
\begin{align}
\bm{x}_t = \bm{x}_{0} + \int_{0}^{t} \bm{h}_{t} dt + \int_{0}^{t} d\bm{w}_t,\label{ddm-diffusion}
\end{align}
where $\bm{x}_{0} \sim q(\bm{x}_{0})$ and $\bm{h}_{t}$ can be expressed as follows
\begin{align}
    \int_{0}^{t} \bm{h}_{t} dt + \bm{x}_{0} = 0. \label{ddm-h}
\end{align}
This leads to the conditional probability distribution for $\bm{x}_t$ as follows.
\begin{align}
q(\bm{x}_t | \bm{x}_0) = \mathcal{N}\left(\bm{x}_t; \bm{x}_0+\int_{0}^{t}\bm{h}_{t} + \int_{0}^{t} d\bm{w}_{t}, t\bm{I}\right).
\end{align}
The probability distribution can also be represented as follows.
\begin{align}
\bm{x}_t = \bm{x}_{0} + \int_{0}^{t} \bm{h}_{t} dt + \sqrt{t} \bm{\epsilon}
\end{align}
where $\bm{\epsilon}$ is a standard Gaussian random variable. To recover the original data $\bm{x}_{0}$ from the noisy observation $\bm{x}_t$, we must solve the reverse SDE, which can be described as follows \cite{huang2024decoupled}.
\begin{align}
d\bm{x}_t = \left[f_t \bm{x}_t - \frac{g_t^2}{\beta_t} \bm{\epsilon}_t\right] dt + g_t d\bar{\bm{w}}_t
\end{align}
In this equation, $\bar{\bm{w}}_t$ is also a standard Wiener process, independent of $\bm{w}_t$, and $\bm{\epsilon}_t$ is a standard Gaussian random variable added to $\bm{x}_{0}$ after being scaled by $\alpha_t$ and $\beta_t$. By setting $f_t \bm{x}_t = \bm{h}_{t}$ and $g_t = 1$, the following equation can be obtained.
\begin{align}
q(\bm{x}_{t-\Delta t} | \bm{x}_t, \bm{x}_{0}) = \mathcal{N}\left(\bm{x}_t + \int_{t}^{t-\Delta t} \bm{h}_t dt - \frac{\Delta t}{t} \bm{\epsilon}, \frac{\Delta t (t - \Delta t)}{t} \bm{I}\right).
\end{align}
The above probability distribution can also be represented as follows.
\begin{align}
\bm{x}_{t-\Delta t}= \bm{x}_t + \int_{t}^{t-\Delta t} \bm{h}_t dt - \frac{\Delta t}{t} \bm{\epsilon}+ \sqrt{\frac{\Delta t (t - \Delta t)}{t}} \widetilde{\bm{\epsilon}},\label{ddm-denoise}
\end{align}
where $\widetilde{\bm{\epsilon}}$ is a normal stochastic variable added in the denoising procedure, which is independent on the diffusion procedure. According to Eq.\eqref{ddm-denoise}, to denoise $\bm{x}_t$, only $\bm{h}_{t}$ and $\bm{\epsilon}$ are unknown terms, where $\hat{\bm{\epsilon}}$ is a standard normal Gaussian random variable ensuring that the variance of the denoised data aligns with that of $\bm{x}_{0}$. This quantity is added artificially and does not require prediction. Empirical studies of denoise diffusion models, such as Stable Diffusion and DALL·E, demonstrate that neural networks must effectively predict $\bm{h}_{t}$ and $\bm{\epsilon}$ for successful image generation. The universal approximation theorem further supports this, indicating that a well-trained neural network can approximate any distribution and mapping, thereby facilitating the denoising of noisy information $\bm{x}_t$.
\begin{remark}
    Since the number of denoising steps in \eqref{ddm-denoise} depends only on $\Delta t$ and $t$, a single denoising step can be performed when $\Delta t = t$. In this case, the SDE-based denoising diffusion can be used for fast inference, achieving a speed comparable to that of traditional NN-based methods, whose inferencing latency and performance are primarily dependent on the size of the NN model.
\end{remark}

\subsection{Relationship Between Diffusion Variable and Noisy Symbol}\label{sec4-2}
Although DMs are typically employed for data generation tasks, regardless of what specific form of the DM, during data generation, the NN does not predict the state $\bm{x}_{t-\Delta t}$ based on the current state $\bm{x}_{t}$. Instead, it predicts the Gaussian noise added between $ \bm{x}_{t-\Delta t} $ and $\bm{x}_{t}$. Therefore, the mathematical foundation of DMs is the removal of Gaussian noise from data, which is why DMs are mathematically referred to as denoising diffusion probabilistic models \cite{ho2020denoising}. The empirical success of DMs in generation tasks demonstrates their effectiveness in eliminating Gaussian noise added to the data. According to \eqref{ddm-denoise}, DMs do not need to denoise from completely random noise at $t=1$. They can initiate the denoising process from any $\bm{x}_{t}$ where $t\in (0,1] $. In fact, during the training procedure of DMs used for data generation, the timestep $t$ is randomly selected, and the DM is trained to predict the Gaussian noise added between $ \bm{x}_{t-\Delta t} $ and $\bm{x}_{t}$ \cite{ho2020denoising,avrahami2022blended}. Therefore, based on the noise content in the data requiring denoising, an appropriate $t$ can be chosen to start the denoising process, thereby recovering clean data.

Moreover, according to \eqref{ddm-diffusion}, as $t$ approaches 1, the proportion of the original data component $\bm{x}_{0}$ in $\bm{x}_{t}$ decreases, while the proportion of the noise component $\bm{\epsilon}$ increases. Conversely, as $t$ approaches 0, the proportion of $\bm{x}_{0}$ in $\bm{x}_{t}$ increases. Therefore, for data with small noise, selecting a large $t$ to start denoising may result in valuable information being discarded as noise. On the other hand, for data with large noise, initiating denoising with a small $t$ may leave residual noise unremoved. This leads to the conclusion that, unlike traditional applications of DMs for content generation, when using DMs for noise reduction, it is crucial to determine how to match the appropriate timestep $t$ according to the characteristics of the data needing denoising. Inspired by information theory, we propose using the SNR as the metric to align the characteristics of the denoised data with the timestep in the DM. This approach is justified because the objective of the DM is to extract the $\bm{x}_{0}$ component from $\bm{x}_{t}$. Similarly, in signal detection, we aim to extract the $ \bm{H}\bm{s} $ component from the received signal $ \bm{r} $. Therefore, it is necessary to find a $t$ such that the proportion of the $\bm{x}_{t}$ component in $\bm{x}_{t}$ matches the proportion of the $ \bm{H}\bm{s}$ component in $\bm{r}$. The $t$ can be calculated as Theorem~\ref{theorem-1} and Remark~\ref{remark-2}.
\begin{theorem}\label{theorem-1}
    Assuming the $\mathbb{E}_{\bm{x}_{0}\sim q(\bm{x}_{0})}\left[\|\bm{x}_{0}\|^{2}\right]=\gamma$, for any given $\mathbb{E}_{\bm{s}\sim\mathcal{S}}\left[\|\bm{Hs}\|^2\right]$ and noise density $\sigma^{2}$ of noisy channel, when 
    \begin{align}
    t = \frac{2+\frac{\mathbb{E}_{\bm{s}\sim\mathcal{S}}\left[\|\bm{Hs}\|^2\right]}{\sigma^2\gamma}-\sqrt{\frac{\mathbb{E}_{\bm{s}\sim\mathcal{S}}\left[\|\bm{Hs}\|^2\right]}{\sigma^2\gamma}\left(\frac{\mathbb{E}_{\bm{s}\sim\mathcal{S}}\left[\|\bm{Hs}\|^2\right]}{\sigma^2\gamma}+4\right)}}{4}\label{t-calculation-1}
    \end{align} 
    following equation can be obtained.
    \begin{align}
        \frac{\mathbb{E}_{\bm{s}\sim\mathcal{S}}\left[\|\bm{Hs}\|^2\right]}{\mathbb{E}_{\bm{n}\sim\mathcal{N}(0,\sigma^2 \bm{I})}\left[\|\bm{n}\|^2\right]}=\frac{\mathbb{E}_{\bm{x}_{0}\sim q(\bm{x}_{0})}\left[\|(1-t)\bm{x}_{0}\|^{2}\right]}{\mathbb{E}_{\bm{\epsilon}\sim \mathcal{N}(0,\bm{I})}\left[\|\sqrt{t}\bm{\epsilon}\|^2\right]}.\label{t-calculation}
    \end{align}
\end{theorem}
\begin{proof}
First, we prove that there is always a $t$ such that \eqref{t-calculation-1} can be held. The part within the root of \eqref{t-calculation-1} is 
\begin{align}
    \frac{\mathbb{E}_{\bm{s}\sim\mathcal{S}}\left[\|\bm{Hs}\|^2\right]}{\sigma^2\gamma}\left(\frac{\mathbb{E}_{\bm{s}\sim\mathcal{S}}\left[\|\bm{Hs}\|^2\right]}{\sigma^2\gamma}+4\right)
\end{align}
Since $\mathbb{E}_{\bm{s}\sim\mathcal{S}}\left[\|\bm{Hs}\|^2\right]$ is the average power of the symbol and is a number greater than zero, the above formula is obviously greater than 0, which means that its square root exists, that is, there exists $t$ to satisfy formula \eqref{t-calculation-1}.

The expected values in the numerator on the right term of \eqref{t-calculation} can be calculated as follows.
\begin{align*}
    \mathbb{E}_{\bm{x}_{0}\sim q(\bm{x}_{0})}\left[\|(1-t)\bm{x}_{0}\|^{2}\right]&=(1-t)^{2}\mathbb{E}_{\bm{x}_{0}\sim q(\bm{x}_{0})}\left[\|\bm{x}_{0}\|^{2}\right],\\
    &=(1-t)^{2}\gamma.
\end{align*}
Similarly, the expected values in the denominator on the right term of \eqref{t-calculation} are:
\begin{align*}
    \mathbb{E}_{\bm{\epsilon}\sim \mathcal{N}(0,\bm{I})}\left[\|\sqrt{t}\bm{\epsilon}\|^2\right]&=t\mathbb{E}_{\bm{\epsilon}\sim \mathcal{N}(0,\bm{I})}\left[\|\bm{\epsilon}\|^2\right],\\
    &=t\left(\mathbb{D}_{\bm{\epsilon}\sim \mathcal{N}(0,\bm{I})}\left[\bm{\epsilon}\right]-\mathbb{E}_{\bm{\epsilon}\sim \mathcal{N}(0,\bm{I})}\left[\bm{\epsilon}\right]^{2}\right)\\
    &=t.
\end{align*}
Therefore, the \eqref{t-calculation} can be rewritten as follows.
\begin{align*}
    \frac{\mathbb{E}_{\bm{s}\sim\mathcal{S}}\left[\|\bm{Hs}\|^2\right]}{\sigma^2\gamma}=\frac{(1-t)^{2}\gamma}{t}.
\end{align*}
Obviously, the above equation is a quadratic equation about $t$, and in order to ensure that $t\in (0,1] $, the Theorem~\ref{theorem-1} can be proved.
\end{proof}
\begin{remark}\label{remark-2}
    In the training procedure, since the purpose of denoise is to solve $\bm{Hs}$, thus the $\bm{x}_{0}$ is set equal to $\bm{Hs}$. Therefore, the $t$ can be simplified as follows.
    \begin{align}
        t = \frac{2\sigma^2+1-\sqrt{1+4\sigma^2}}{4\sigma^2}.\label{calculate-t}
    \end{align}
\end{remark}

According to Theorem~\ref{theorem-1}, there always exists a suitable timestep $t$ such that the proportion of noise components in $\bm{x}_t$ matches that of in the received signal $\bm{r}$. However, during the training process, only $\bm{x}_0$ is known, and $\bm{x}_t$ must be computed using \eqref{ddm-diffusion} to train the diffusion model (DM). This implies that the trained DM is proficient at extracting features from the data distribution of $\bm{x}_t$ for $t \in (0,1]$. Nevertheless, the distribution of the received signal $\bm{r}$ may differ from that of $\bm{x}_t$, and directly inputting $\bm{r}$ into the DM can lead to an out-of-distribution scenario. This mismatch results in a significant decline in the DM's denoising performance. Therefore, it is essential to transform the data distribution of $\bm{r}$ to be consistent with that of $\bm{x}_t$.

Generally, altering the distribution of arbitrary data types is extremely challenging. However, in signal detection, due to the specific distribution characteristics of signals and noise, this alignment can be achieved through simple mathematical scaling. Since the goal of denoising is to recover $\bm{H}\bm{s}$, we have $\bm{x}_0 = \bm{H}\bm{s}$. Additionally, because $\bm{x}_t$ is a linear combination of $\bm{x}_0$ and zero-mean Gaussian noise $\boldsymbol{\epsilon}$, and $\bm{r}$ is also a linear combination of $\bm{H}\bm{s}$ and zero-mean noise $\bm{n}$, the form of their distributions remains unchanged under linear combinations—only their statistical characteristics are modified. Therefore, by ensuring that the average norms of $\bm{r}$ and $\bm{x}_t$ are equal, we can align their distributions. Since $\bm{x}_t$ is computed based on Theorem \ref{theorem-1}, we opt to apply a linear transformation to $\bm{r}$ to match its distribution with that of $\bm{x}_t$. This approach also guarantees that the $\bm{x}_0$ estimated by the DM corresponds to the required $\bm{H}\bm{s}$ without additional computations. The linear scaling of $\bm{r}$ is formalized in Theorem~\ref{theorem-2}.

\begin{theorem}\label{theorem-2}
    By scaling $\bm{r}$ with a factor of $\alpha$, the scaled $\alpha \bm{r}$ is transferred into the distribution of $\bm{x}_{t}$, where the $\alpha$ is as follows.
    \begin{align}
        \alpha = \sqrt{\frac{(1-t)^{2}\mathbb{E}_{s\sim\mathcal{S}}[\|\bm{Hs}\|^2]+t}{\mathbb{E}_{\bm{s}\sim\mathcal{S}}\left[\|\bm{Hs}\|^2\right]+\sigma^2}}
    \end{align}
\end{theorem}
\begin{proof}
The expected value of $\mathbb{E}_{\bm{s}\sim \mathcal{S}, \bm{n}\sim \mathcal{N}(0,\sigma^{2}\bm{I})}\left[\|\alpha \bm{r}\|^{2}\right]$ is as follows.
    \begin{align*}
        \mathbb{E}_{\bm{s}\sim \mathcal{S}, \bm{n}\sim \mathcal{N}(0,\sigma^{2}\bm{I})}\left[\|\alpha \bm{r}\|^{2}\right]&=\mathbb{E}_{\bm{s}\sim \mathcal{S}, \bm{n}\sim \mathcal{N}(0,\sigma^{2}\bm{I})}\left[\|\alpha (\bm{s}+\bm{n})\|^{2}\right],\\
        &=\alpha^2 \left(\mathbb{E}_{\bm{s}\sim \mathcal{S}}\left[\| \bm{s}\|^{2}\right]+\mathbb{E}_{\bm{n}\sim \mathcal{N}(0,\sigma^{2}\bm{I})}\left[\|\bm{n}\|^{2}\right]\right).
    \end{align*}
Moreover, the expected value of $\mathbb{E}_{\bm{x}_{0}\sim q(\bm{x}_{0})}\left[\|\bm{x}_{t}\|^{2}\right]$ is as follows.
\begin{align*}
    \mathbb{E}_{\bm{x}_{0}\sim q(\bm{x}_{0})}\left[\|\bm{x}_{t}\|^{2}\right]&=\mathbb{E}_{\bm{x}_{0}\sim q(\bm{x}_{0})}\left[\|(1-t)\bm{x}_{0}+\sqrt{t}\bm{\epsilon}\|^{2}\right],\\
    &=(1-t)^{2}\mathbb{E}_{\bm{x}_{0}\sim q(\bm{x}_{0})}\left[\|\bm{x}_{0}\|\right]+t\mathbb{E}_{\bm{\epsilon}\sim\mathcal{N}(0,\bm{I})}\left[\|\bm{\epsilon}\|^{2}\right],\\
    &=(1-t)^{2}\mathbb{E}_{\bm{x}_{0}\sim q(\bm{x}_{0})}\left[\|\bm{x}_{0}\|\right] +t
\end{align*}
Then by solving the following equation the value of $\alpha$ can be obtained as Theorem~\ref{theorem-2}.
\begin{align*}
    \mathbb{E}_{\bm{x}_{0}\sim q(\bm{x}_{0})}\left[\|\bm{x}_{t}\|^{2}\right]= \mathbb{E}_{\bm{s}\sim \mathcal{S}, \bm{n}\sim \mathcal{N}(0,\sigma^{2}\bm{I})}\left[\|\alpha \bm{r}\|^{2}\right].
\end{align*}
It should be noted that although the above formula can guarantee that the obtained $\alpha$ can scale the expectation of $\bm{r}$ to be consistent with $\bm{x}_{t}$, it cannot directly guarantee that the distribution of $\bm{r}$ and $\bm{x}_{t}$ is consistent. However, according to Remark~\ref{remark-2} because there is $\bm{x}_{0} = \bm{s}$, we can think that $\bm{r}$ and $\bm{x}_{t}$ obey the same distribution form, and $\alpha$ ensures the same expectation between the two. For the adjustment of variance, according to \eqref{ddm-denoise}, the amplitude of noise added in the denoising process can be adjusted.
\end{proof}

\begin{figure}
    \centering
    \includegraphics[width=1\linewidth]{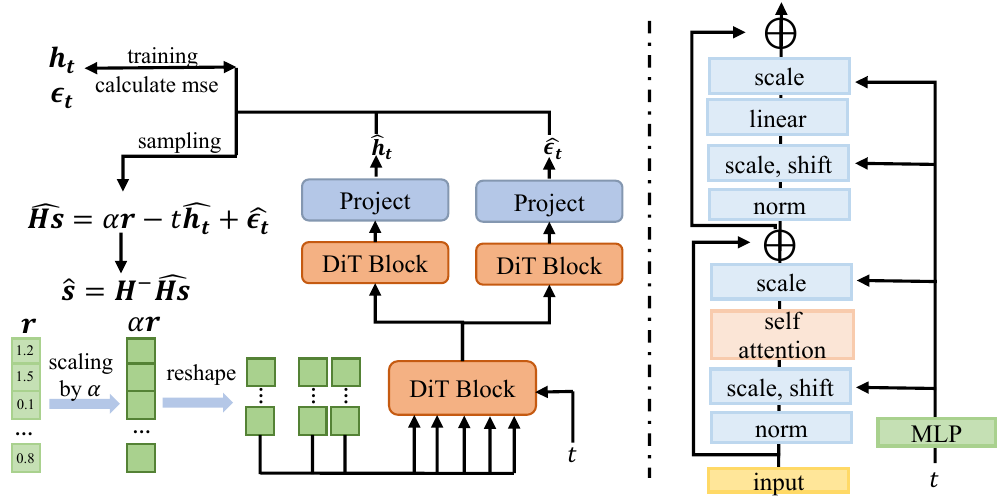}
    \caption{The illustration of the inferencing and training procedure of the proposed DM on the left, and the details of the NN architecture on the right.}
    \label{fig-2}
\end{figure}

\section{DM-Based signal detection Method}\label{sec5}
Based on the analysis in Section~\ref{sec4}, as is shown in Fig.~\ref{fig-1} we propose a DM-based signal detection method is proposed as follows.
\begin{align}
&\bm{\hat{h}}_{t}, \bm{\hat{\epsilon}}_{t} = \bm{\epsilon}_{\bm{\theta}}(\alpha\bm{r}, t),\label{pred-h-e}\\
&\bm{\hat{\mathbf{H}\mathbf{s}}} = \alpha\bm{r} - t\bm{\hat{h}}_{t} - \bm{\hat{\epsilon}}_{t},\\
&\bm{\hat{s}} = \bm{H}^{-1}\bm{\hat{\mathbf{H}\mathbf{s}}},
\end{align}
where $\bm{\epsilon}_{\bm{\theta}}$ denotes the DM implemented as a neural network with trainable parameters $\bm{\theta}$, and $t$ and $\alpha$ are calculated according to Theorem~\ref{theorem-1} and Theorem~\ref{theorem-2}, respectively. In the denoising procedure, both $\bm{h}_{t}$ and $\bm{\epsilon}_{t}$ are unknown variables that need to be estimated. 

According to \cite{huang2024decoupled}, it is essential to utilize a NN to predict these values. Specifically, we adopt the DiT architecture to design a NN capable of predicting both $\bm{h}_{t}$ and $\bm{\epsilon}$ as is shown in Fig.~\ref{fig-2}. The input to the NN is the received signal $\bm{r}$, while the denoising timestep $t$ serves as a conditioning input. Since DiT is fundamentally based on the Transformer architecture, tokenization of the input data is necessary. Specifically, we partition the received signal $\bm{r}$ into $N_{\text{tok}}$ equal-length vectors, referred to as tokens, which serve as the input to the DiT model as follows.
\begin{align}
\bm{r} \rightarrow [\bm{r}_1, \bm{r}_2, \dots, \bm{r}_{N_{\text{tok}}}],
\end{align}
where each $\bm{r}_i$ is a tokenized segment of $\bm{r}$. The self-attention mechanisms are then employed to extract features from the data as follows.
\begin{align}
\text{Attention}(\bm{Q}, \bm{K}, \bm{V}) &= \text{softmax}\left(\frac{\bm{Q}\bm{K}^\mathrm{T}}{\sqrt{d}}\right)\bm{V},\label{attention}\\
\bm{Q} &= \bm{W}_{Q} \bm{r},\\
\bm{K} &= \bm{W}_{K} \bm{r},\\
\bm{V} &= \bm{W}_{V} \bm{r},
\end{align}
where $\bm{W}_{Q}$, $\bm{W}_{K}$, and $\bm{W}_{V}$ are trainable weight matrices, and $d$ is the dimensionality of the queries and keys. Although self-attention effectively extracts data features, controllable denoising is required to restore the original signal, as indicated by \eqref{ddm-denoise} and Theorem~\ref{theorem-1}. Unlike traditional Transformers that use cross-attention or additional tokens for conditioning, DiT incorporates the timestep $t$ by modulating the Transformer's output characteristics through scalar scaling and shifting. This modulation is described as follows.
\begin{align}
\bm{z} &= \bm{z} \odot (1 + \zeta) + \kappa,\\
\zeta &= \bm{W}_{\zeta} t, \quad \kappa = \bm{W}_{\kappa} t,
\end{align}
where $\bm{z}$ is the output from the self-attention mechanism, $\odot$ denotes element-wise multiplication, and $\zeta$ and $\kappa$ are scaling and shifting vectors calculated by trainable matrices $\bm{W}_{\zeta}$ and $\bm{W}_{\kappa}$, respectively, applied to $t$. After stacking $N$ DiT blocks to fully extract data features, we proceed to predict $\bm{h}_{t}$ and $\bm{\epsilon}$. Additional DiT blocks are employed to separately predict the features of $\bm{h}_{t}$ and $\bm{\epsilon}$, projecting them as $\bm{\hat{h}}_{t}$ and $\bm{\hat{\epsilon}}_{t}$ with the same dimensions as $\bm{h}_{t}$ and $\bm{\epsilon}_{t}$. Following \cite{huang2024decoupled}, the training loss of the DiT is defined as follows.
\begin{align}
\mathcal{L} = \mathbb{E}\left[\|\bm{\hat{h}}_{t} - \bm{h}_{t}\|^{2} + \|\bm{\hat{\epsilon}}_{t} - \bm{\epsilon}_{t}\|^{2}\right].
\end{align}

\begin{theorem}
    The inferencing computation complexity of the proposed DM-based method is $\mathcal{O}(|r|^2)$, where $|\bm{r}|$ is the cardinality of the $\bm{r}$.
\end{theorem}
\begin{proof}
    Since the proposed DM-based method uses the Transformer as the lead backbone, computational complexity is mainly determined by the Transformer network. According to \cite{vaswani2017attention}, the computational complexity of the Transformer is $\mathcal{O}(n^2)$, where $n$ is the number of tokens. Since we use the received signal $\bm{r}$ patchify as tokens, the number of tokens is at most the same as the dimension of $\bm{r}$ that is $|\bm{r}|$. At the same time, according to \eqref{ddm-denoise}, DM can implement one-step denoising, so the computational complexity is $\mathcal{O}(|\bm{r}|^2)$.
\end{proof}

\section{Simulation Results}
\subsection{Implementation}
In the simulation, we investigate a signal detection system with various $N_{r}$, specifically $N_r = 4$, $8$, and $16$. For the simulation source data type, we generate random transmission symbols following a Bernoulli distribution where each transmission symbol is randomly selected from the $\mathcal{S}$ with equal probability. 

We implemented our framework using the PyTorch library, ensuring a professional and efficient setup. The training process employs the AdamW optimizer with an initial learning rate of $5 \times 10^{-5}$ and a weight decay of $1 \times 10^{-4}$. A custom warm-up learning rate scheduler is applied, where the learning rate linearly increases during the first 250 iterations to stabilize the training process. After the warm-up phase, the learning rate gradually decreases to improve convergence. An exponential moving average (EMA) of the model parameters is maintained throughout training, with a decay rate of 0.9996. EMA updates commence after 1,000 steps and occur every 8 steps thereafter, enhancing the model's generalization capabilities. For the experiments using random signal sequences as input, we trained the model on a single NVIDIA A100 GPU with a batch size of 1,024 over 50,000 training steps. The inputs are noisy sequences processed by a Transformer module with a depth of 28 layers, enabling the model to capture complex patterns effectively. For the experiments utilizing the MNIST dataset, training was conducted on a single NVIDIA RTX 4090 GPU with a batch size of 128. The inputs are noisy image signals, also processed by a 28-layer Transformer module, allowing the model to learn semantic information from the images efficiently. We set the gradient accumulation steps to 1 and did not use mixed-precision training. Training progress is logged every 20 iterations, and model checkpoints along with sample results are saved every 2,000 steps. The results are periodically saved to a designated folder for subsequent analysis.

\subsection{Comparisons with Other Methods}

To assess the performance of our proposed model, the following baseline methods are used to be compared.
\begin{itemize}
    \item \textbf{ML}: The method that has long been regarded as the optimal signal detection method, which exhaustively enumerates all possible symbol combinations and selects the one with the highest likelihood as the estimated transmission symbols.
    \item \textbf{MMSE}: A classical linear signal detection method, which inverts the received signal by applying the channel noise with the regularized inverse of the channel matrix.
    \item \textbf{FC-DNN}: An end-to-end signal detection method based on multi-layer perception (MLP) architecture proposed in \cite{liu2023exploiting}.
    \item \textbf{DnCNN}: A deep convolutional neural network (CNN) based method, where the DnCNN that is originally designed for image denoising is used for signal detection in \cite{dncnn}.
    \item \textbf{SigT}: A Transformer-based signal detection method that employs self-attention mechanisms to capture global dependencies in the data in \cite{sigt}.
\end{itemize}

\begin{figure}[h]
    \centering
    \subfigure[SER on 4$\times$4 MIMO with different SNR.]
    {
       \centering
       \includegraphics[width=0.3\columnwidth]{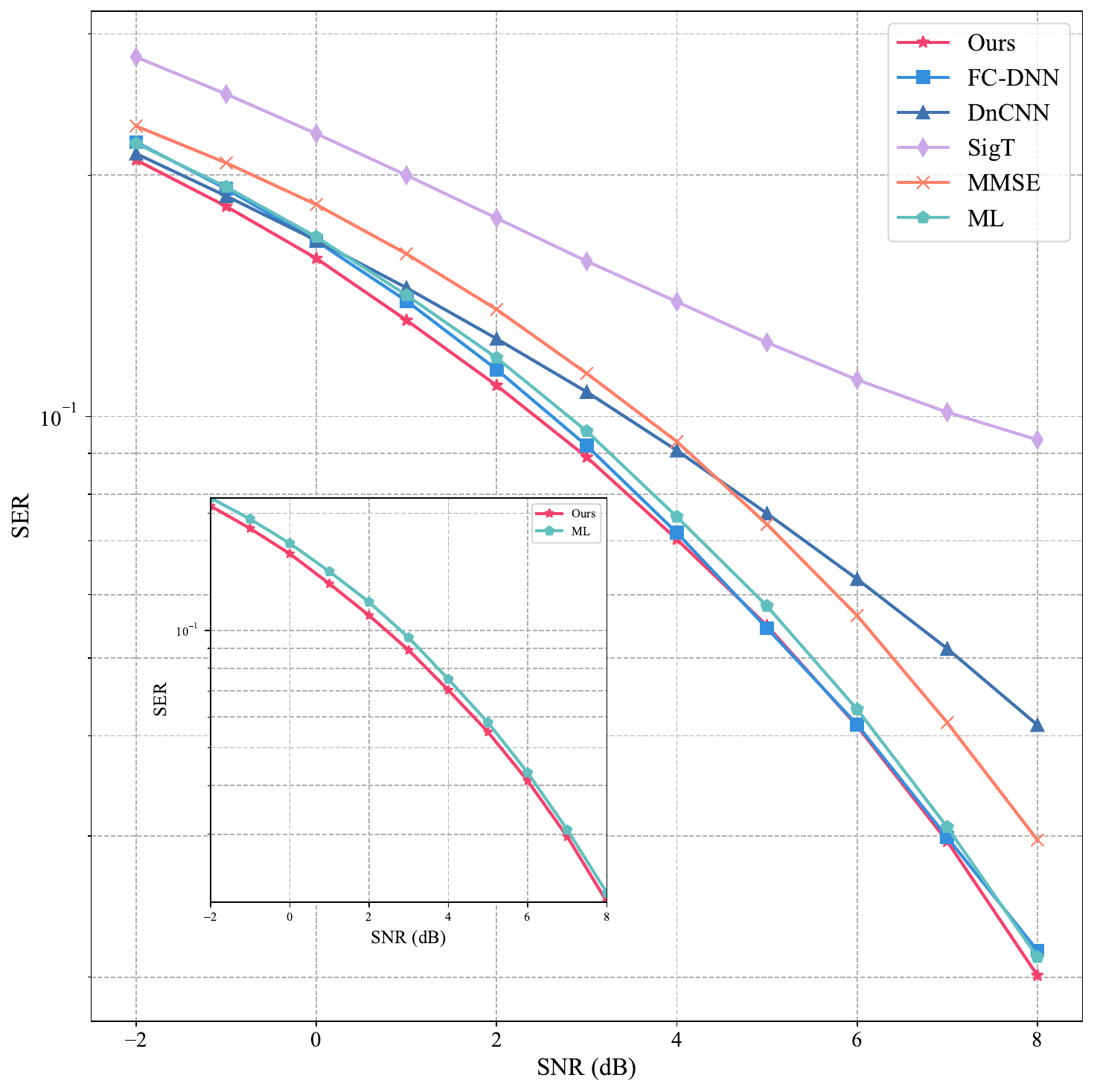}\label{fig-random-4-bpsk}
    }
    \subfigure[SER on 8$\times$8 MIMO with different SNR.]
    {
       \centering
       \includegraphics[width=0.3\columnwidth]{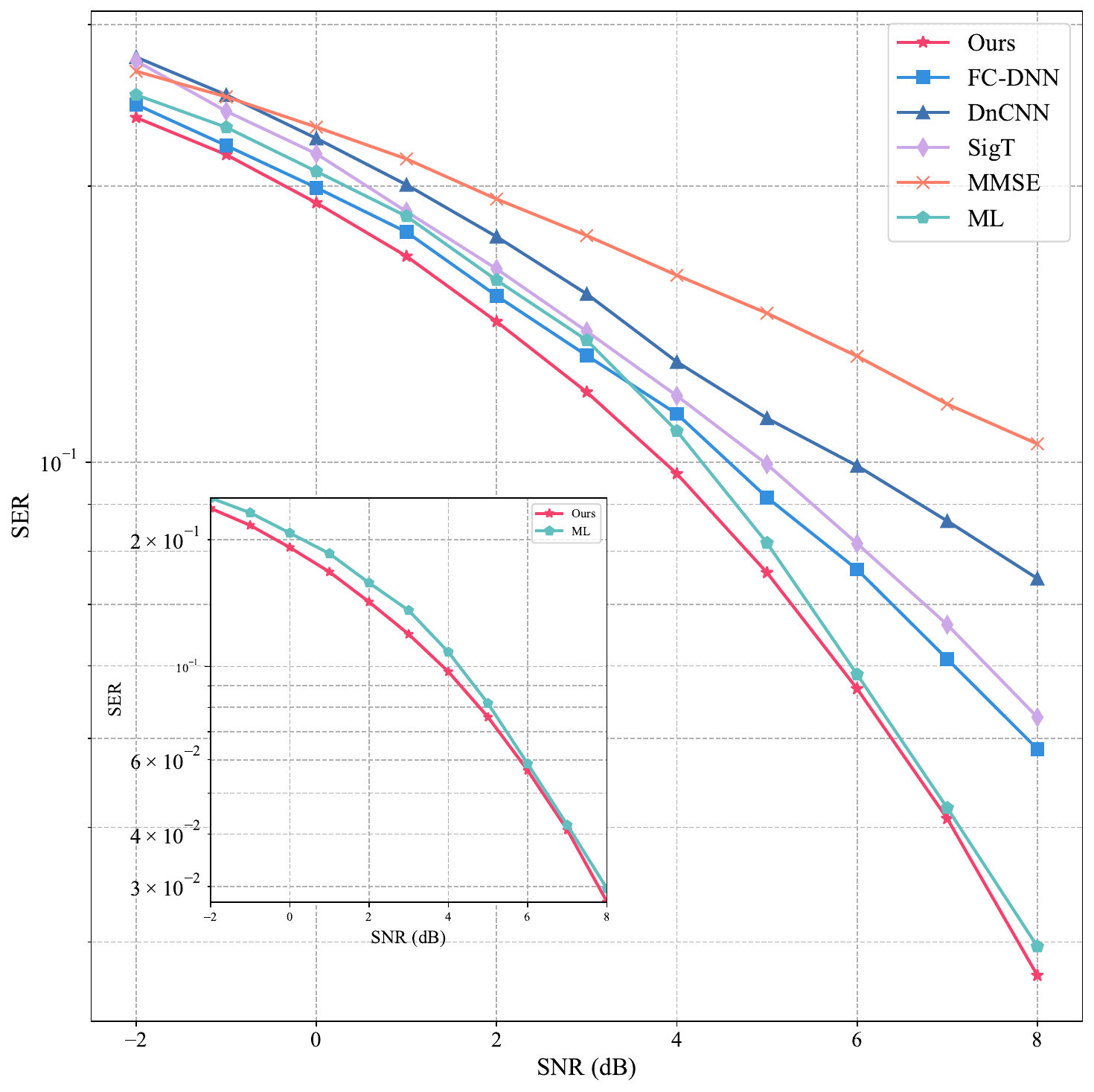}\label{fig-random-8-bpsk}
    }
    \subfigure[SER on 16$\times$16 MIMO with different SNR.]
    {
       \centering
       \includegraphics[width=0.3\columnwidth]{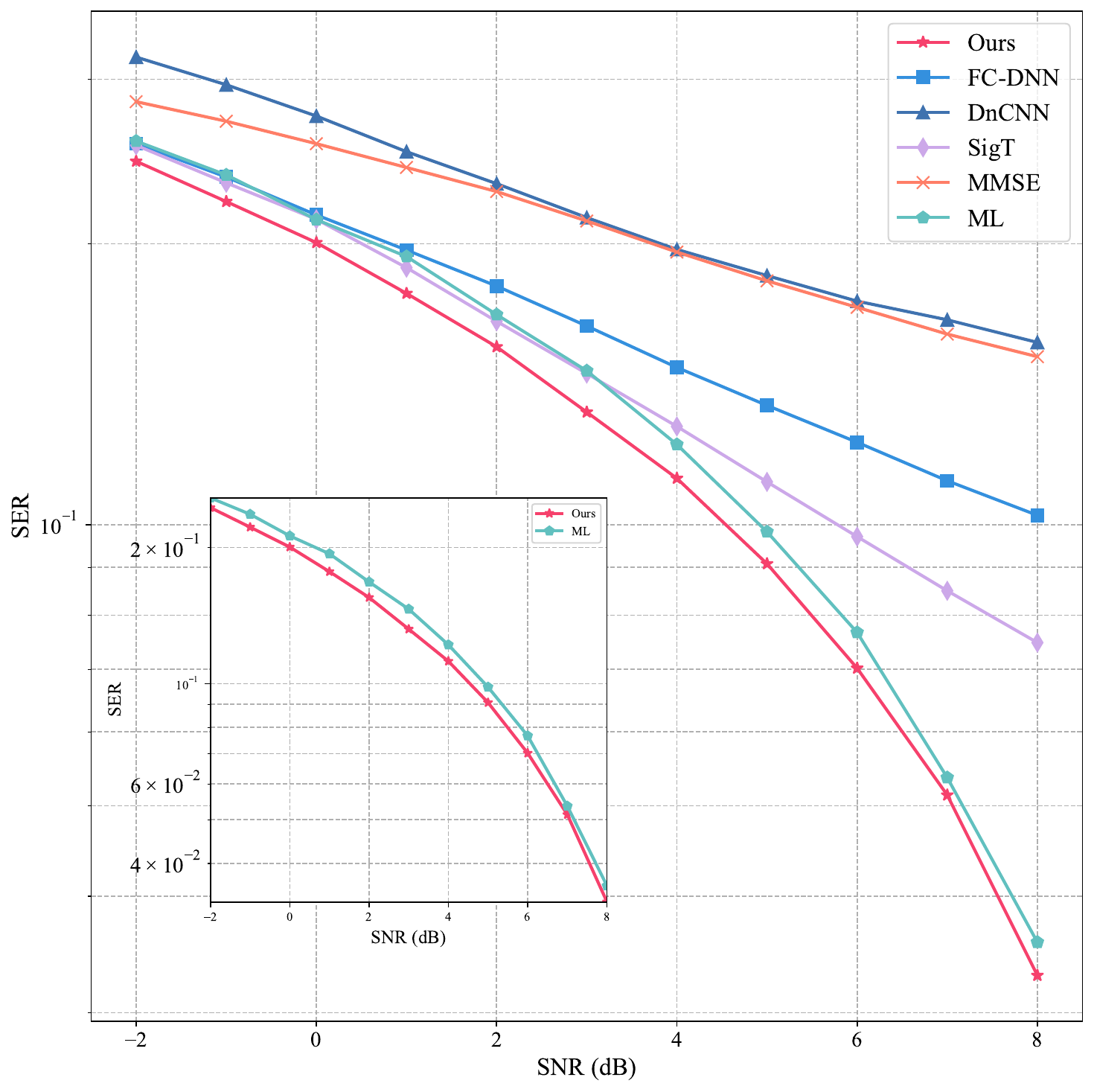}\label{fig-random-16-bpsk}
    }
    \caption{SER on different SNR and $N_r$ for different signal detection methods with BPSK modulation.}
    \label{fig-random-bpsk}
\end{figure}

\begin{figure}[h]
    \centering
    \subfigure[SER on 4$\times$4 MIMO with different SNR.]
    {
       \centering
       \includegraphics[width=0.3\columnwidth]{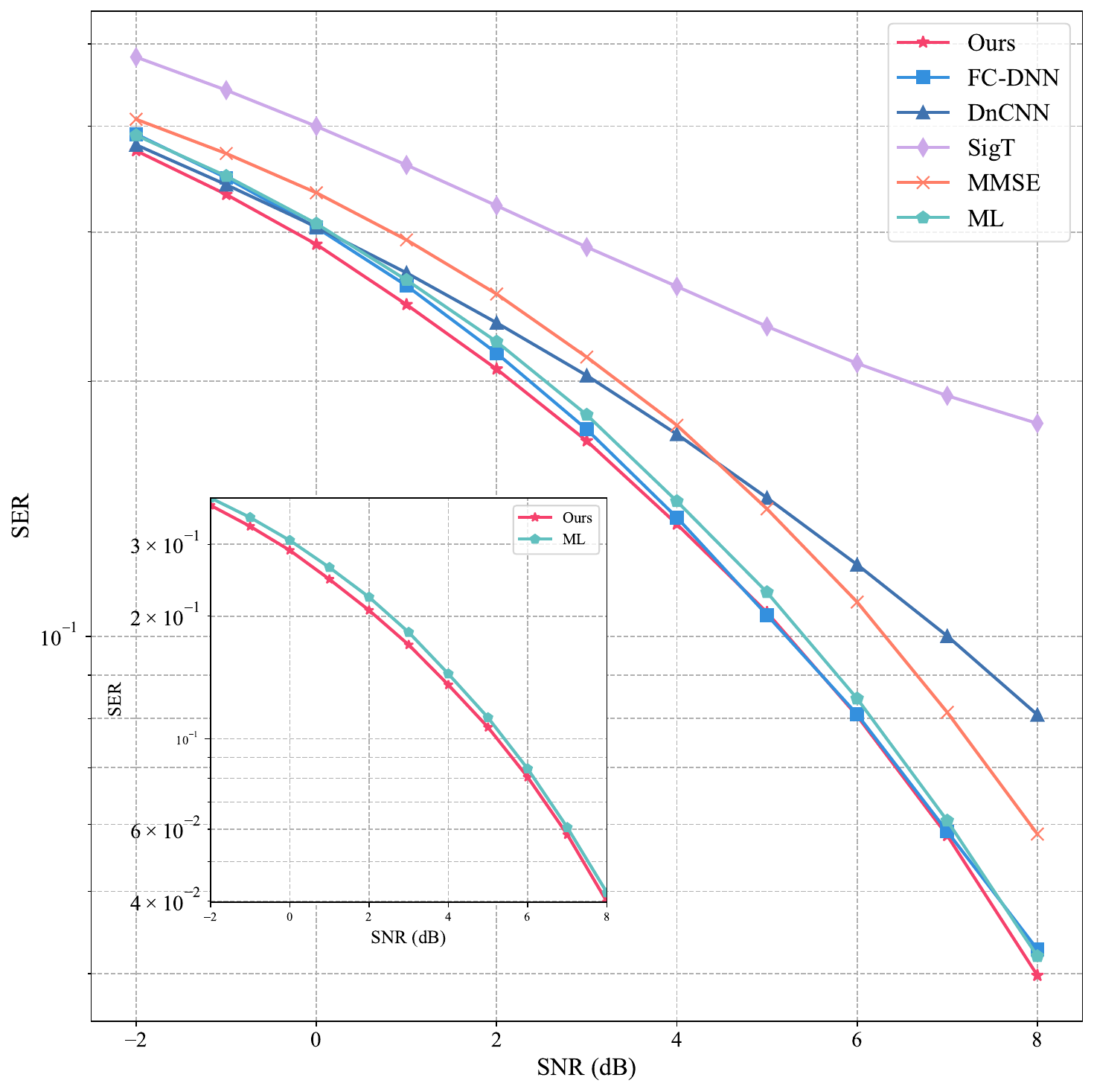}\label{fig-random-4-qam}
    }
    \subfigure[SER on 8$\times$8 MIMO with different SNR.]
    {
       \centering
       \includegraphics[width=0.3\columnwidth]{random_res/NS_4_qpsk.pdf}\label{fig-random-8-qam}
    }
    \subfigure[SER on 16$\times$16 MIMO with different SNR.]
    {
       \centering
       \includegraphics[width=0.3\columnwidth]{random_res/NS_4_qpsk.pdf}\label{fig-random-16-qam}
    }
    \caption{SER on different SNR and $N_r$ for different signal detection methods with 4-QAM modulation.}
    \label{fig-random-qam}
\end{figure}

Fig.~\ref{fig-random-bpsk} present a comparison between the proposed DM-based signal detection method and all baseline methods under binary phase-shift keying (BPSK) modulation. As shown in the figures, the DM-based signal detection algorithm consistently outperforms the ML estimation—the traditional optimal signal detection method—across all SNR and $N_r$ settings. This demonstrates that the proposed DM-based intelligent signal detection framework effectively extends traditional signal detection methodologies. Based on this theory, it is possible to design new algorithms that surpass the performance of classical optimal receiving methods. Moreover, the DM-based algorithm exhibits a steeper decline in SER with increasing SNR compared to the ML method. Although both methods show similar SER trends—where SER decreases as SNR increases—the SER for the DM-based method remains consistently lower than that of ML. In contrast, other deep learning-based methods initially achieve a reduction in SER as SNR increases, but their performance plateaus at higher SNRs. This behavior is commonly observed in traditional NN-based signal detection algorithms, such as those using MLP, CNN, and Transformer architectures. This observation highlights that the primary limitation of NN-based signal detection methods lies in the underlying signal detection theory rather than in the network architecture itself. Additionally, it is noteworthy that the DM-based signal detection algorithm was trained only once under a specific $N_r$ and did not require fine-tuning across different SNR settings. This result validates Theorems~\ref{theorem-1} and \ref{theorem-2}, which show that by identifying the appropriate timestep $t$ and scaling factor $\alpha$ corresponding to the received signal's SNR through mathematical analysis, the impact of OOD data on NN-based signal detection performance can be effectively mitigated.

In Fig.~\ref{fig-random-qam}, we further compare the SER of different signal detection methods under 4-quadrature amplitude modulation (4-QAM). Since 4-QAM can be regarded as a combination of two BPSK signals, the downward trend of SER for different signal detection methods in a 4-QAM modulation system is similar to that in BPSK. However, because the BER in dual signals accumulates, the DM-based signal detection method, having a lower BER in single signals, achieves a significantly lower SER in the 4-QAM system compared to other algorithms.

\subsection{Performance Analysis for High-Order QAM Modulation and Large-Scale MIMO}

\begin{figure}[h]
    \centering
    \subfigure[SER on 16-QAM.]
    {
       \centering
       \includegraphics[width=0.45\columnwidth]{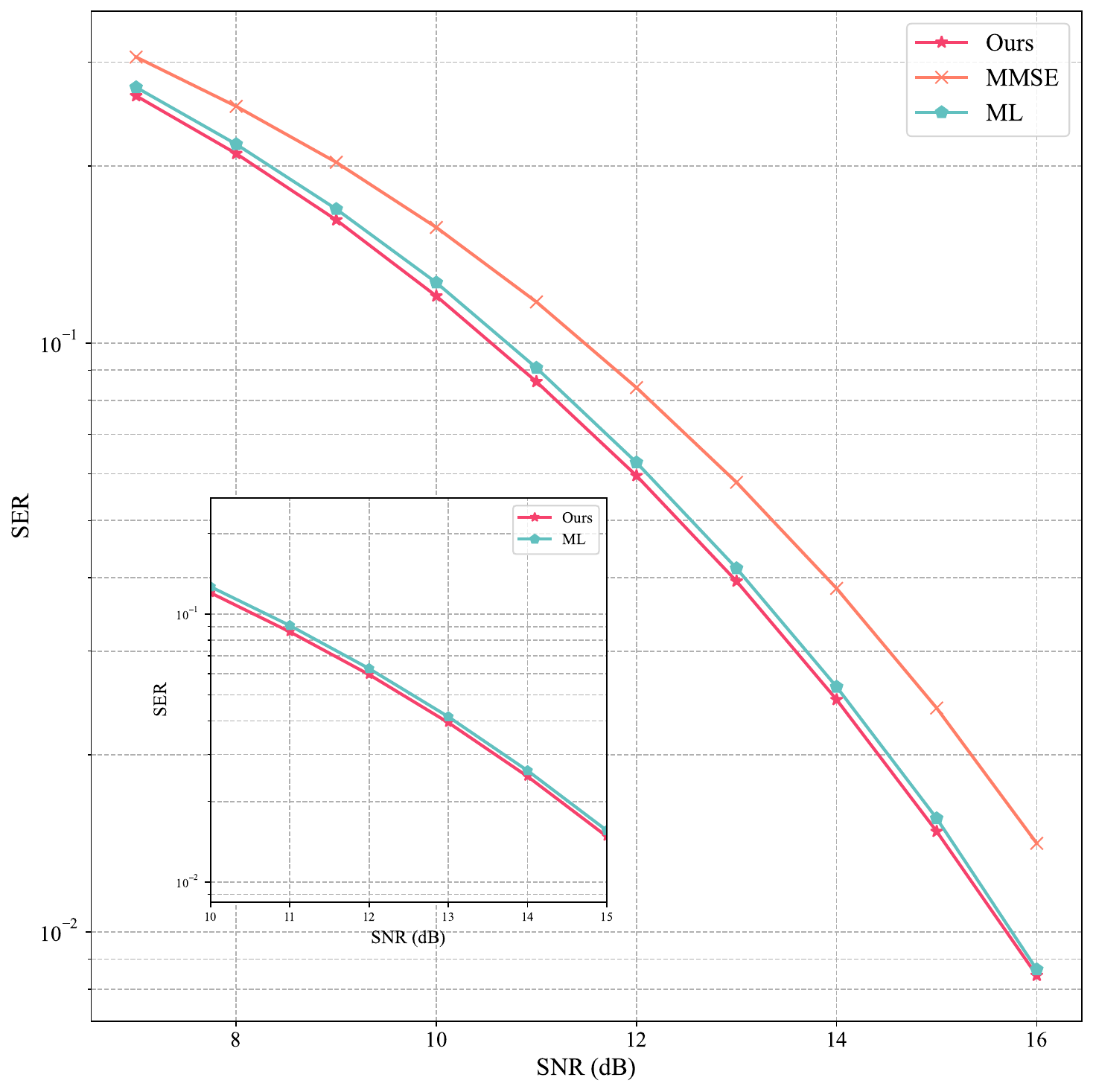}\label{fig-16qam}
    }
    \subfigure[SER on 64-QAM.]
    {
       \centering
       \includegraphics[width=0.45\columnwidth]{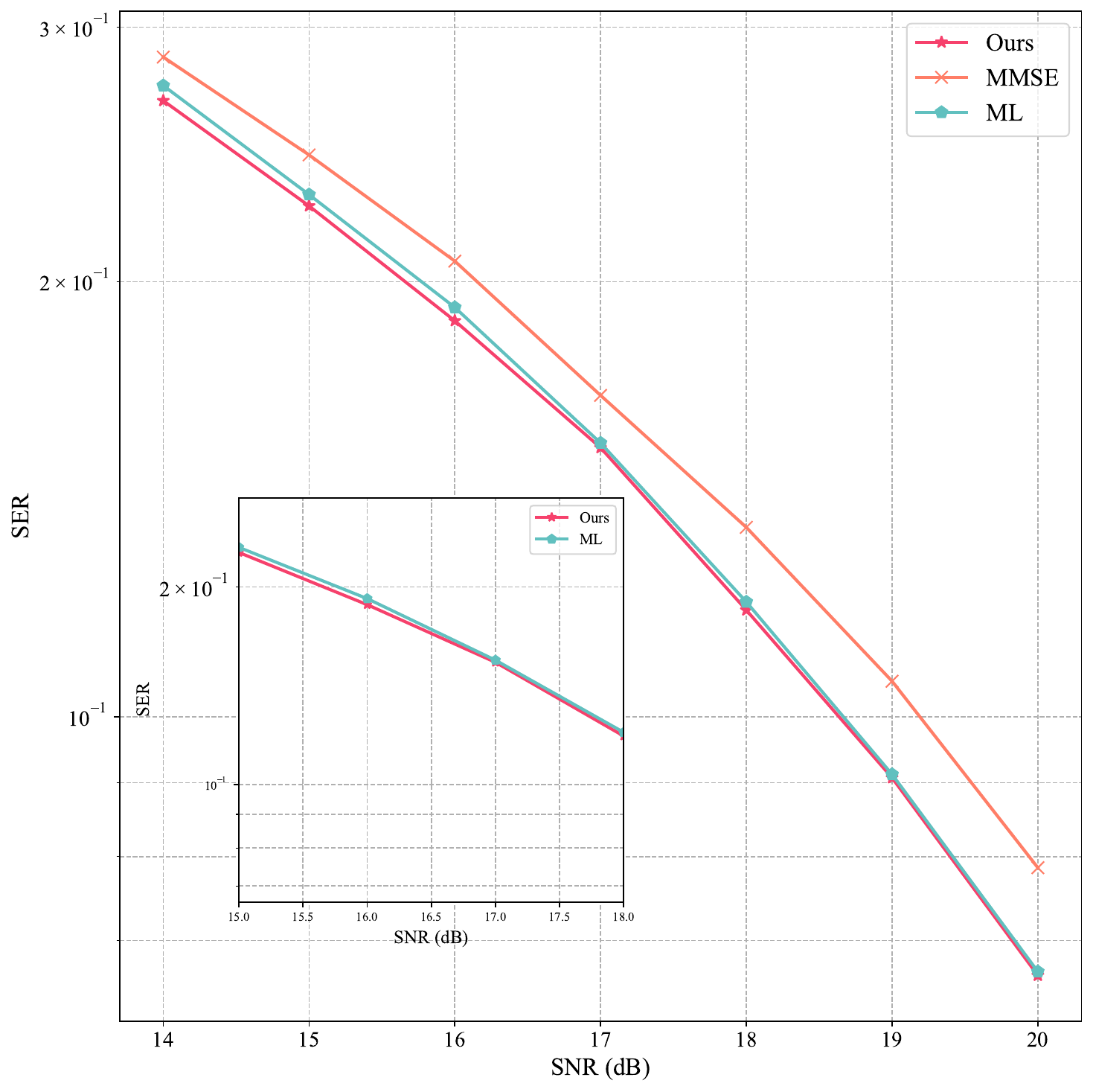}\label{fig-64qam}
    }
    \caption{SER on 4$\times$4 MIMO with high-order QAM.}
    \label{fig-QAM}
\end{figure}

To further evaluate the performance of the proposed DM-based signal detection method in high-order modulation systems, we conducted simulations focusing on systems with $N_{r} = 4$ receiving antennas. Specifically, we compared the performance of our method under high-order QAM schemes with baseline algorithms, including traditional linear and optimal signal detection methods. In this study, we considered 16-QAM, 64-QAM, and 256-QAM modulation schemes, which are representative of high-order modulation used in advanced communication systems such as 6G networks. High-order QAM increases spectral efficiency by encoding more bits per symbol but reduces the distance between constellation points. This reduction makes the system more susceptible to noise, resulting in a higher SER under the same SNR conditions compared to lower-order QAM. Fig.~\ref{fig-QAM} illustrates the SER performance of different signal detection methods as a function of SNR for the 16-QAM and 64-QAM modulation schemes. Due to the computational complexity of maximum likelihood (ML) estimation for 256-QAM requiring $256^{N_r}$ combinations—it is impractical to include ML results in our comparisons for this modulation order. Notably, for 16-QAM and 64-QAM, the DM-based method also achieves a lower SER than the ML estimation method, which is traditionally considered the optimal signal detection technique. These results demonstrate that the proposed DM-based signal detection method is widely applicable to various modulation schemes, including high-order QAM. The superior performance indicates that our method effectively mitigates the challenges associated with the reduced distance between modulation symbols in high-order QAM. By enhancing the reliability and error-free transmission rate of communication systems, the DM-based method addresses a critical requirement for modern networks. Improving the performance of high-order modulation methods is particularly significant for 6G networks, which demand high data rates and efficient spectrum utilization to support data-intensive services.

\subsection{Numerical Analysis on Theorem~\ref{theorem-1} and Theorem~\ref{theorem-2}}

\begin{table*}[h]
    \centering
    \setlength{\tabcolsep}{1.2mm}{} 
    \renewcommand{\arraystretch}{1}
    \newcolumntype{P}[1]{>{\centering\arraybackslash}p{#1}}
    \caption{The $t$ and $\alpha$ calculated by Theorem~\ref{theorem-1} and \ref{theorem-2}}
    \resizebox{0.9\linewidth}{!}{
\begin{tabular}{c|c|c|c|c|c|c|c|c|c|c|c}
\toprule \hline
SNR (dB) & -5  & -3 & -1 &  1  &  3  &  5 &  7  &  9 &  11  &  13 & 15 \\ \hline
$t$  &  0.6275  &  0.5587  &  0.4854  &  0.4083  &  0.3337  &  0.2620  &  0.1978  &  0.1441  &   0.1017  &   0.0697 & 0.0467 \\ \hline
$\alpha$ &  0.3725  &  0.4413  &  0.5146  &  0.5917  &  0.6663  &  0.7380  &  0.8022  &  0.8559  &   0.8983  &   0.9303  & 0.9533\\ \hline
\bottomrule
\end{tabular}
}
    \vspace{-12pt}
    \label{tab-t-alpha}
\end{table*}

\begin{figure}[h]
    \centering
    \includegraphics[width=0.4\linewidth]{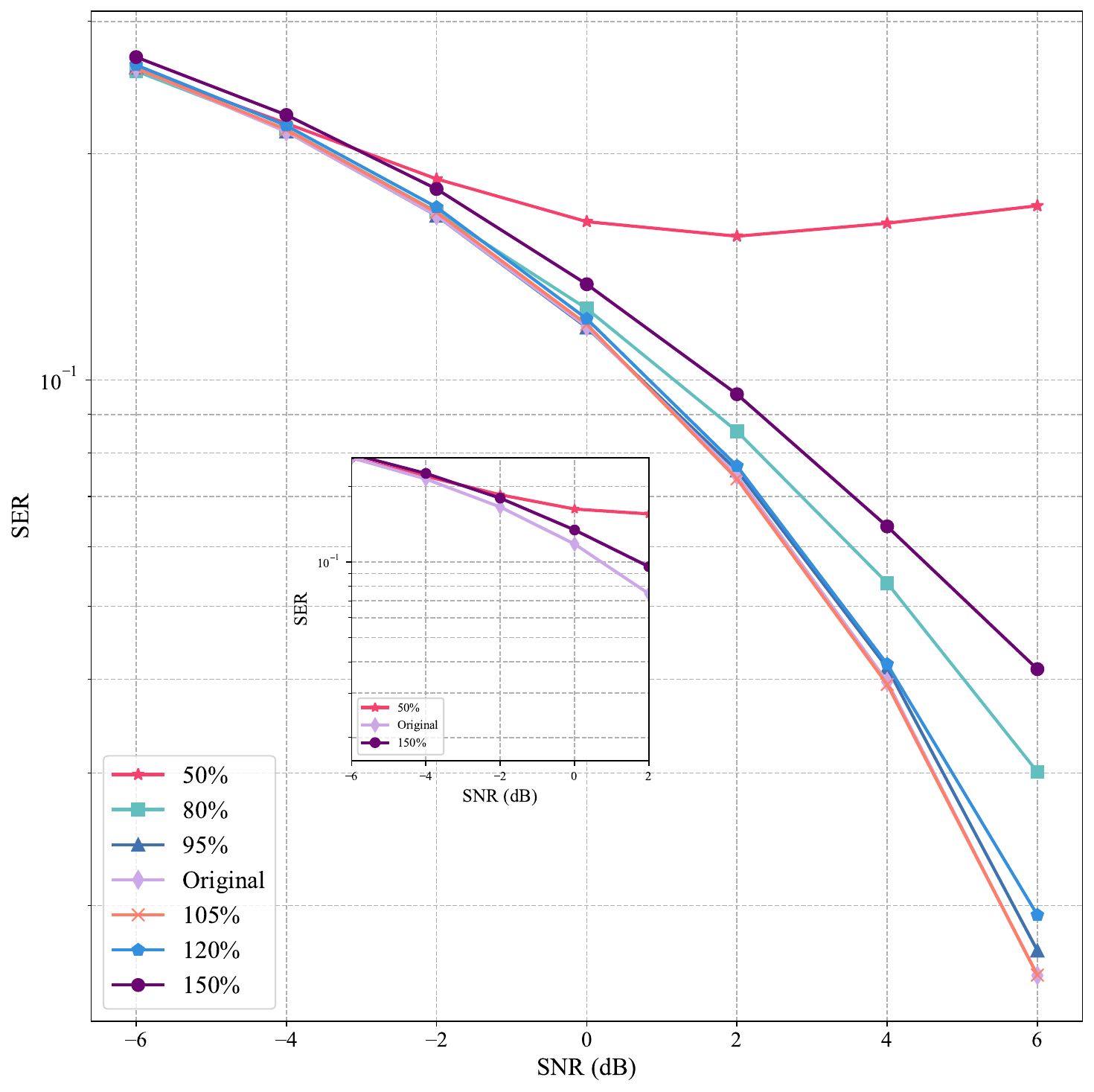}
    \caption{SER on different $\alpha$.}
    \label{fig-alpha}
\end{figure}

In this section, we utilize numerical simulations to validate the correctness of the timestep $t$ and scaling factor $\alpha$ computed in Theorems \ref{theorem-1} and \ref{theorem-2}. Table~\ref{tab-t-alpha} presents the theoretical values of $t$ and $\alpha$ calculated using these theorems. It can be observed that as the SNR increases, the value of $t$ gradually decreases. This trend occurs because, as $t$ approaches zero, the proportion of the original signal component $\bm{x}_0$ in $\bm{x}_t$ increases, which is consistent with the conditions of higher SNR where the noise component is relatively smaller. Furthermore, as the SNR increases, the value of $\alpha$ gradually increases and approaches 1. This behavior is due to the fact that at higher SNRs, the received signal $\bm{r}$ becomes increasingly similar to $\bm{x}_0$ in the DM, leading $\alpha$ to approach 1. Conversely, when the SNR is low, $\alpha$ is significantly less than 1, indicating the necessity to appropriately scale $\bm{r}$ in low-SNR conditions. This necessity is also reflected in Fig.~\ref{fig-alpha}, which illustrates the impact of scaling on the performance at different SNR levels. These observations confirm the correctness of the theoretical calculations of $t$ and $\alpha$ as per Theorems \ref{theorem-1} and \ref{theorem-2}. The numerical results validate the effectiveness of the proposed method in adapting to varying SNR conditions, supporting the theoretical framework established in this paper.

\begin{figure}[h]
    \centering
    \includegraphics[width=0.4\linewidth]{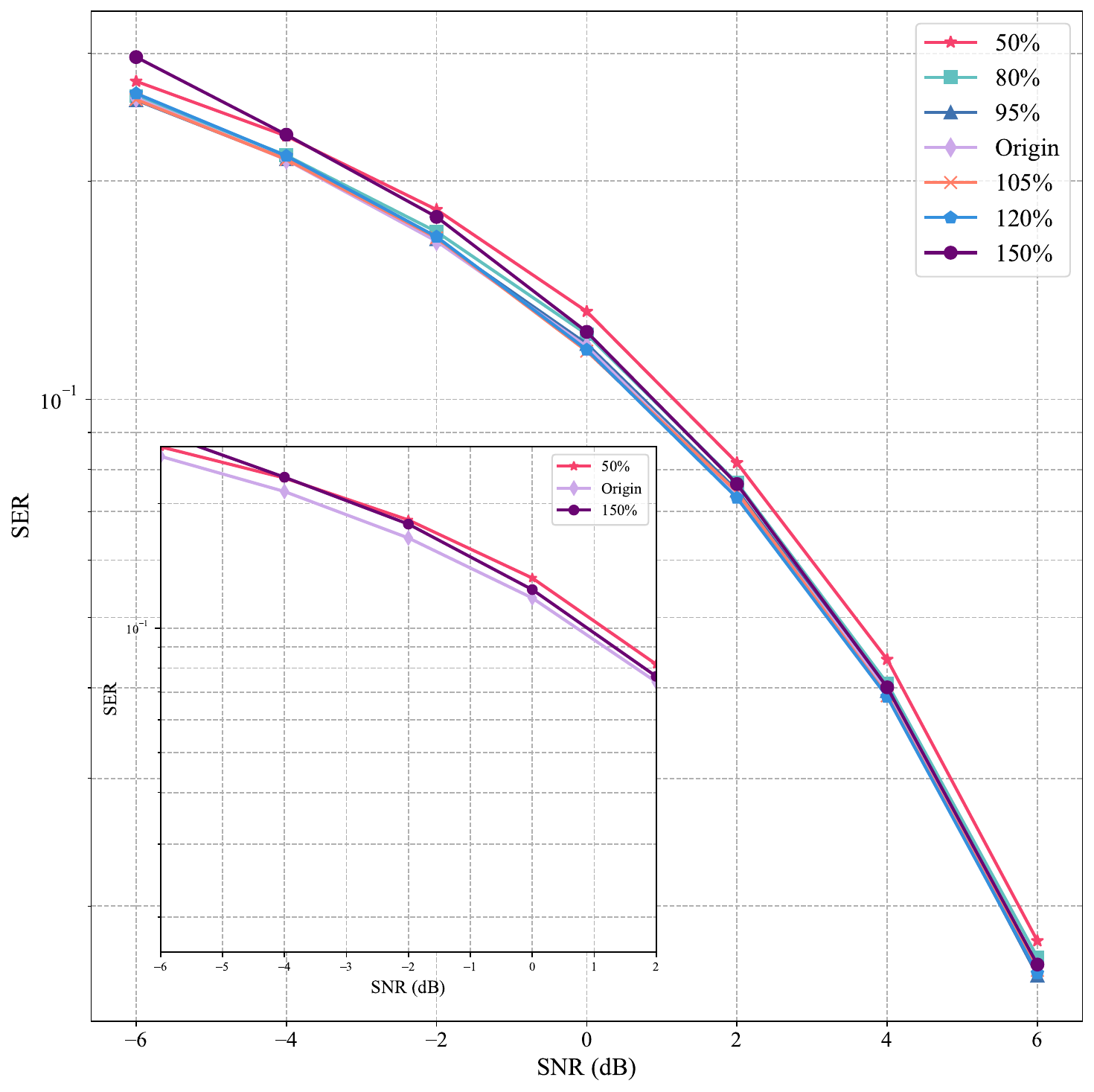}
    \caption{SER on different $t$.}
    \label{fig-t}
\end{figure}

In Fig.~\ref{fig-t}, we compare the SER of the proposed DM-based signal detection method using the original timestep $t$ calculated according to Theorem~\ref{theorem-1} with the SER obtained when $t$ is increased and decreased by 5\%, 20\%, and 50\%, respectively, under different SNRs. The experiments are conducted with $N_r = 8$ using BPSK modulation. It is evident that, under any SNR, the DM-based signal detection method utilizing the $t$ calculated from Theorem~\ref{theorem-1} significantly outperforms the methods with adjusted $t$ values. As the value of $t$ deviates from the calculated $t$, the SER of signal detection increases. This performance degradation occurs because an excessively large $t$ leads the DM to assume that the signal contains more noise components than it actually does, resulting in the erroneous removal of useful signal information. Conversely, an excessively small $t$ causes insufficient noise removal, leaving residual noise in the received signal. These observations confirm the correctness of Theorem~\ref{theorem-1} in numerical experiments. However, it is noteworthy that when the actual $t$ deviates only slightly from the value calculated by Theorem~\ref{theorem-1}, the performance does not decrease drastically. This demonstrates the robustness and practicality of our proposed algorithm, as it allows for small errors in estimating the channel noise power—an important consideration in real-world communication systems.

Figure~\ref{fig-alpha} illustrates the effect of different scaling factors $\alpha$ on the performance of the DM-based signal detection method. We compare the SER of the original $\alpha$ calculated according to Theorem~\ref{theorem-2} with the SER obtained when $\alpha$ is increased and decreased by 5\%, 20\%, and 50\%, respectively, under different SNRs, with $N_r = 8$ using BPSK modulation. It is observed that deviations in $\alpha$ calculation significantly reduce the performance of the DM-based signal detection method compared to changes in the $t$ value. Particularly, when the actual $\alpha$ deviates by 50\% from the theoretical value, the SER of the DM-based method remains higher than 10\%, rendering the method almost ineffective. This performance fluctuation occurs because a significant deviation in $\alpha$ leads to a substantial mismatch between the distribution of the received signal $\bm{r}$ input to the DM and the corresponding $\bm{x}_t$, making $\bm{r}$ out-of-distribution (OOD) data. Consequently, the performance of the DM-based signal detection method degrades. Nonetheless, it is evident that the $\alpha$ calculated using Theorem~\ref{theorem-2} consistently achieves the best performance, confirming the accuracy and effectiveness of Theorem~\ref{theorem-2} in numerical experiments. Furthermore, the calculation of $\alpha$ exhibits a certain degree of robustness; when the deviation is not more than 5\%, its impact on SER is insignificant. This robustness facilitates the application of DM-based signal detection methods in practical systems, where exact parameter estimation may be challenging.

\begin{figure}
    \centering
    \includegraphics[width=0.4\linewidth]{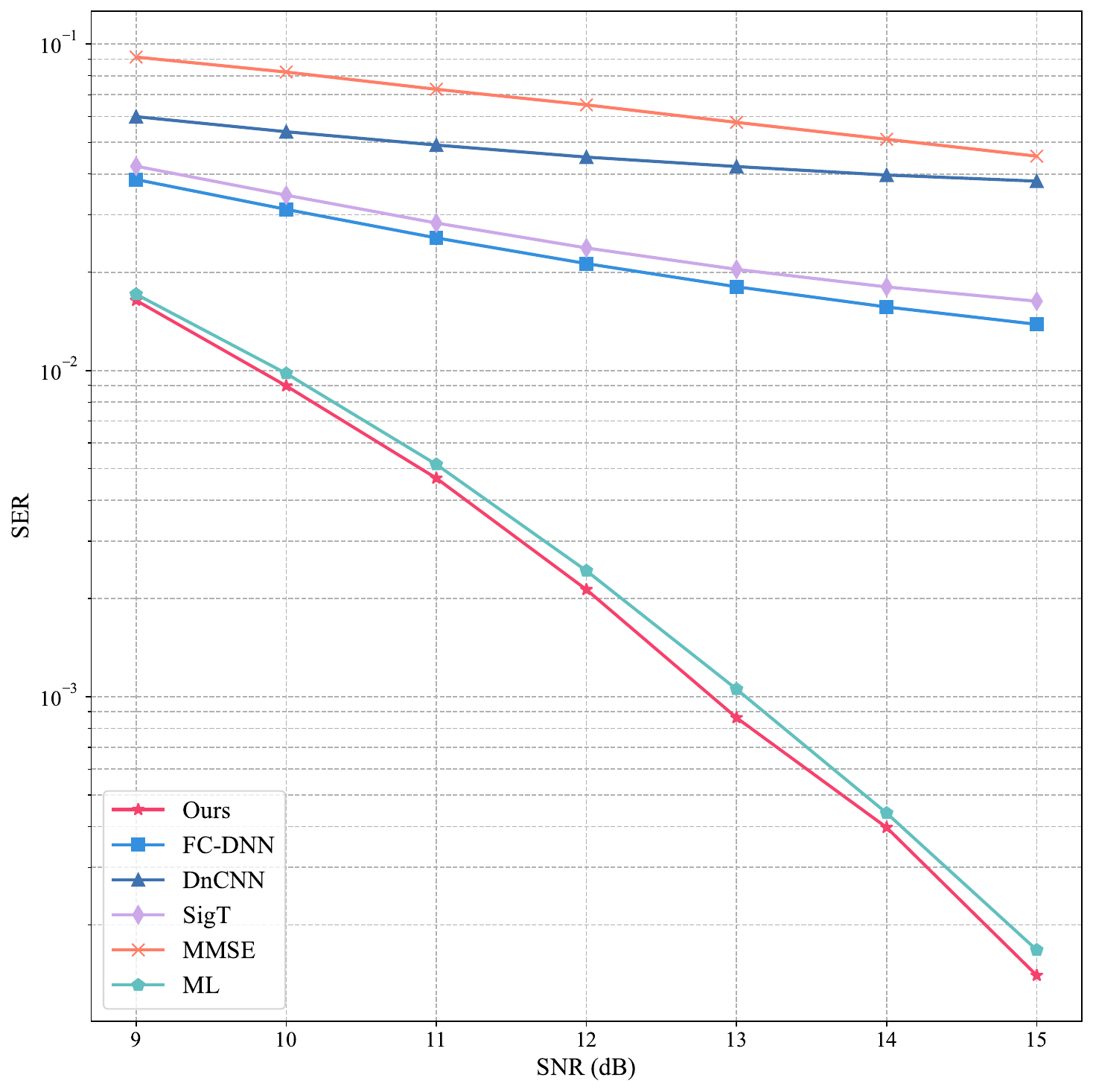}
    \caption{SER on high SNR.}
    \label{fig-high-snr}
\end{figure}

According to Table~\ref{tab-t-alpha}, it is evident that when the SNR is large, the value of the timestep $t$ becomes very close to zero. In traditional DM training methods, where $t$ is uniformly and randomly selected from the range $(0, 1]$, the DM may not be adequately trained for high-SNR conditions corresponding to small $t$ values. To address this limitation, in Fig.~\ref{fig-high-snr}, we retrained the DM with $t$ selected from the range $(0, 0.2]$, ensuring sufficient training for high-SNR scenarios. We observe that at high SNRs, the ratio of noise to the received signal becomes very small. Traditional NN-based signal detection methods, constrained by the theoretical frameworks guiding their training, struggle to effectively distinguish the noise component from the signal under these conditions. As a result, beyond a certain SNR threshold, the SER of these NN-based methods hardly decreases with further increases in SNR. In contrast, traditional signal detection methods such as MMSE and ML estimation not only exhibit decreasing SER with increasing SNR but also demonstrate an accelerated rate of decline. Consequently, at high SNRs, the performance of traditional NN-based methods becomes significantly inferior to that of the ML method and may even be worse than MMSE. However, the signal detection theory established in this paper provides a theoretical advancement over traditional approaches. The DM-based signal detection method we propose maintains superior performance compared to ML even at high SNRs. Moreover, the rate of decline in SER for the DM-based method increases with increasing SNR, attributed to the focused training of the DM for small $t$ values. These results fully illustrate the effectiveness of our proposed method.

\subsection{Ablation Study}

\begin{figure}[h]
    \centering
    \includegraphics[width=0.4\linewidth]{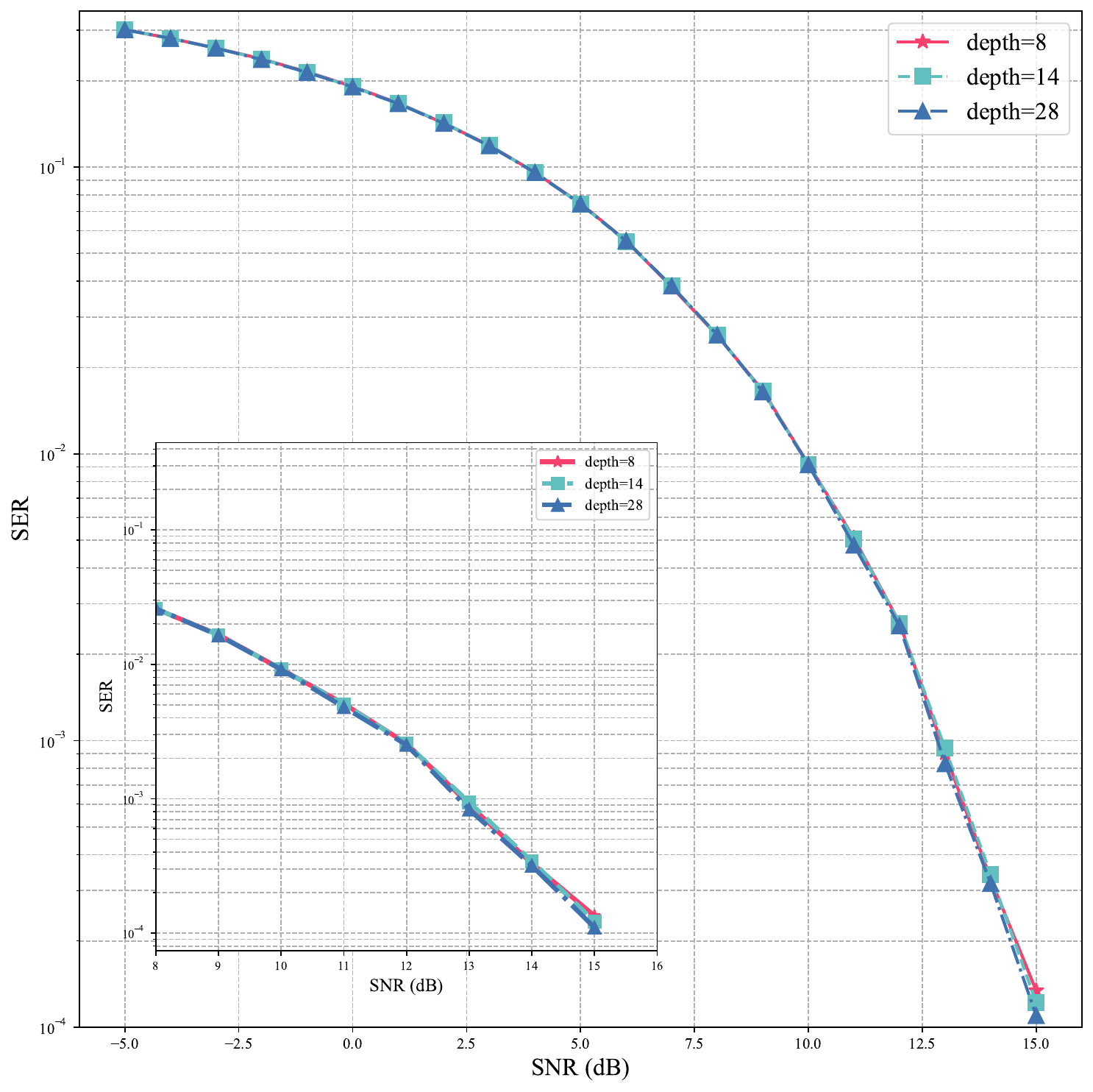}
    \caption{SER on different NN depths.}
    \label{fig-depth}
\end{figure}

\begin{figure}[h]
    \centering
    \includegraphics[width=0.4\linewidth]{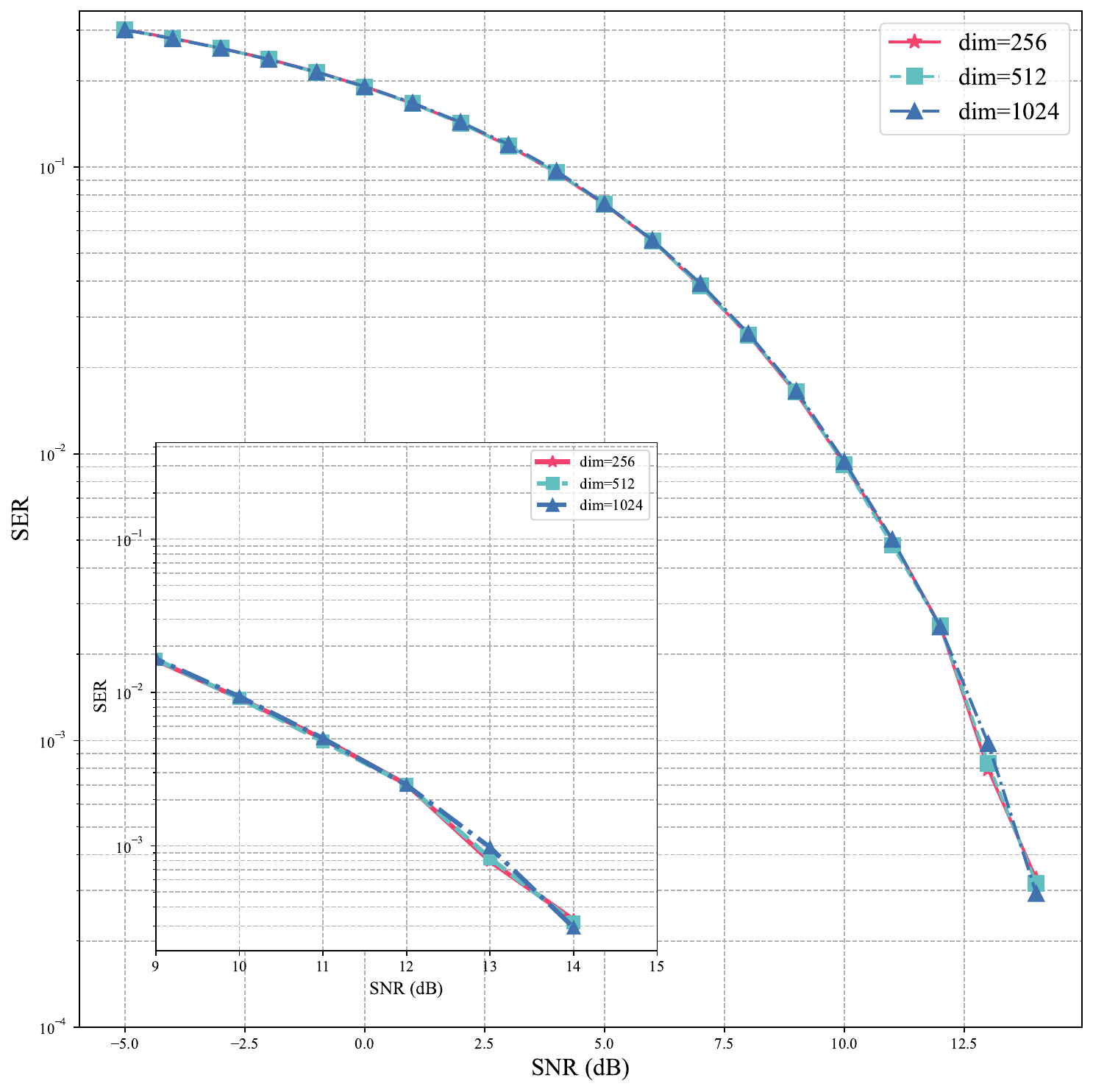}
    \caption{SER on different NN dimensions.}
    \label{fig-dim}
\end{figure}

Traditionally, the depth and width of NNs significantly influence their feature extraction capabilities, which in turn affect the performance of NN-based signal detection methods trained using traditional approaches. To verify the versatility of our proposed DM-based intelligent signal detection theory, we conducted ablation experiments on the depth and dimension of the DiT. Specifically, we varied the number of DiT blocks as depths and the length of the $\bm{Q}$, $\bm{V}$, and $\bm{K}$ vectors in its self-attention mechanism in \eqref{attention} as dimensions. The experimental results are presented in Fig.~\ref{fig-depth} and Fig.~\ref{fig-dim} with $N_r=8$ and BPSK modulation. It is evident from the figures that altering the depth and dimension of the DiT does not significantly impact the performance of signal detection. This observation indicates that our proposed intelligent signal detection theory and DM-based signal detection method exhibit strong generalization across different NN architectures. Furthermore, it demonstrates that the performance gains achieved are not attributable to the enhanced feature extraction capabilities of the NN, but rather to the advanced theoretical framework we have introduced.

\section{Conclusion}
In this paper, we have established a pioneering intelligent signal transmission theory based on the SDE forms of the DM, which mathematically demonstrates that additive Gaussian noise can be effectively reduced using the DM. Moreover, we have established a mathematical correlation between the SNR of the received signal and the timestep of the DM. This correlation enables a trained DM to be utilized for signal detection under any SNR through linear scaling. Based on this new intelligent signal transmission theory, we have proposed a novel signal receiving method whose performance surpasses the theoretical upper limit of traditional optimal signal detection methods. Numerical simulations have validated the correctness and advancements of our proposed theory. By applying the DM-based signal detection method proposed in this paper to 6G networks, the reliability and effectiveness of signal transmission can be significantly improved, thereby enhancing the quality of service for hyper reliable \& low-latency communications and immersive communication services such as virtual and augmented reality. This method directly addresses the pressing demands of 6G networks for higher capacity and transmission rates without relying on hardware advancements, which are becoming increasingly limited in the post-Moore's Law era. By effectively erasing communication noise through intelligent signal processing, our approach provides a novel solution to overcome the bottlenecks related to spectral efficiency and computational complexity in current systems. To further apply the established intelligent signal transmission theory to practical communication systems, future research will focus on employing variational autoencoders (VAEs) \cite{mescheder2017adversarial} to handle additive non-Gaussian noise by transforming arbitrary data distributions into latent Gaussian distributions.

\bibliography{ref}
\bibliographystyle{IEEEtran}

\ifCLASSOPTIONcaptionsoff
  \newpage
\fi

\end{document}